\documentclass[screen,sigconf,nonacm]{acmart}

\usepackage{booktabs,colortbl,multicol,multirow}
\usepackage[table]{xcolor}
\newcommand{\graycell}[1]{\cellcolor{gray!10}#1}

\usepackage{float}

\usepackage{url}
\newcommand{\gradcell}[1]{%
  \begingroup
  \pgfmathsetmacro{\val}{#1}%
  \def\cellshade{}%
  \ifdim\val pt>0pt
    \pgfmathtruncatemacro{\shade}{min(80,round(80*\val/0.75))}%
    \xdef\cellshade{\noexpand\cellcolor{poscolor!\shade!white}}%
  \else
    \ifdim\val pt<0pt
    \pgfmathtruncatemacro{\shade}{min(80,round(80*abs(\val)/0.75))}%
      \xdef\cellshade{\noexpand\cellcolor{negcolor!\shade!white}}%
    \fi
  \fi
  \endgroup
  \cellshade #1%
}

\definecolor{negcolor}{HTML}{FF69B4}
\definecolor{poscolor}{HTML}{006400}

\usepackage{multirow}

\definecolor{DarkBlue}{HTML}{00008B}
\definecolor{mscolor}{HTML}{01665e}
\definecolor{nmscolor}{HTML}{bf812d}
\definecolor{lgreen}{HTML}{ccece6}
\definecolor{dolive}{HTML}{308014}
\definecolor{purple}{HTML}{ae017e}
\definecolor{brickred}{HTML}{f03b20}

\newif{\ifhidecomments}
  \hidecommentsfalse 
\ifhidecomments
    \newcommand{\agam}[1]{}
    \newcommand{\claire}[1]{}
    \newcommand{\koustuv}[1]{}
    \newcommand{\eshwar}[1]{}
\else
    \newcommand{\agam}[1]{\textbf{\small\sffamily{\textcolor{teal}{[#1 -- Agam]}}}}
    \newcommand{\claire}[1]{\textbf{\small\sffamily{\textcolor{DarkBlue}{[#1 -- Claire]}}}}
    \newcommand{\koustuv}[1]{\textbf{\small\sffamily{\textcolor{orange}{[#1 -- Koustuv]}}}}
    \newcommand{\eshwar}[1]{\textbf{\small\sffamily{\textcolor{brickred}{[#1 -- Eshwar]}}}}
  \fi

\usepackage{subcaption}
\AtBeginDocument{%
  }

\setcopyright{acmlicensed}
\copyrightyear{2018}
\acmYear{2018}
\acmDOI{XXXXXXX.XXXXXXX}
\acmConference[Conference acronym 'XX]{Make sure to enter the correct
  conference title from your rights confirmation email}{June 03--05,
  2018}{Woodstock, NY}
\acmISBN{978-1-4503-XXXX-X/2018/06}

\begin{document}
\sloppy
\title[First Impressions: How Placement Shapes the Influence of AI Summaries]{First Impressions: How Placement Shapes the Influence of AI Summaries}

\author{Wang Claire}
\orcid{0009-0003-3562-055X}
\affiliation{%
  \institution{University of Illinois Urbana-Champaign}
  \city{Urbana}
  \state{IL}
  \country{USA}
}
\email{claire46@illinois.edu}

\author{Agam Goyal}
\orcid{0009-0009-5989-2887}
\affiliation{%
  \institution{University of Illinois Urbana-Champaign}
  \city{Urbana}
  \state{Illinois}
  \country{USA}}
\email{agamg2@illinois.edu}

\author{Frederick Choi}
\orcid{0000-0002-8818-2456}
\affiliation{%
  \institution{University of Illinois Urbana-Champaign}
  \city{Urbana}
  \state{IL}
  \country{USA}
}
\email{fc20@illinois.edu}

\author{Koustuv Saha}
\orcid{0000-0002-8872-2934}
\affiliation{%
  \institution{University of Illinois Urbana-Champaign}
  \city{Urbana}
  \state{Illinois}
  \country{USA}}
\email{ksaha2@illinois.edu}

\author{Eshwar Chandrasekharan}
\orcid{0000-0002-7473-1418}
\affiliation{%
  \institution{University of Illinois Urbana-Champaign}
  \city{Urbana}
  \state{Illinois}
  \country{USA}}
\email{eshwar@illinois.edu}

\renewcommand{\shortauthors}{Claire et al.}

\begin{abstract}
AI-generated summaries increasingly mediate how people interpret information across platforms, including product reviews on e-commerce sites. Using Amazon's AI summaries as a case study, we conducted a preregistered, randomized experiment (N = 278) comparing how AI summaries and user reviews shaped product perceptions, and how their influence varied with valence and presentation order. We found that both AI summaries and user reviews influenced participants’ opinions, with negative summaries having a larger effect than positive ones. Presentation order was the most important factor: the first source anchored judgment and only user reviews could displace an existing anchor. Although participants reported preferring user reviews, they often underestimated the influence of AI summaries on their judgments. Our findings show how the placement of AI summaries shapes user perception and highlight opportunities to design interfaces that support more deliberate judgments about when to rely on summaries and when to examine the underlying content directly.\end{abstract}



\maketitle

\section{Introduction}

AI summaries have become more prevalent across contexts. They have been integrated into search engines, review sites, and social media sites, on platforms such as Google Reviews, Google Search, Docusign, and Slack. In traditional means of online information access, users seek crowdsourced information, such as search results or reviews, drawing on others' experiences to form their own opinions and make their own decisions. However, the volume of this information can be difficult to parse for a user. Thus, AI summaries provide an immediate utility: instead of picking through every individual search result or review themselves, users can consult these LLM-powered summaries for high-level summaries, trading comprehensiveness for convenience. However, these information systems have also drawn critique for lacking agency, transparency, and intellectual friction~\cite{informationaccess} due to the black-box nature by which these summaries are generated and by which they are presented to the user. 

Though LLMs can help users make more informed decisions, they also have the potential to promote misinformation and disrupt user autonomy in ways of which that users may not even be aware. 
LLMs have been found to amplify biases around gender, race, and other axes of marginalization~\cite{cheng2023marked, park2024generative,williams2023epidemic, yao2025generative, ghaffarzadegan2024generative, park2023generative, zou2024can, serapio2023personality, wang2024incharacter, tu2023characterchat}. 
In the context of AI-assisted decision making, prior work has shown that manipulated AI explanations can sway people to make unfair decisions~\cite{li2024utilizinghumanbehaviormodeling}. 
Therefore, when envisioning AI-assisted decision making, it is imperative to assess and calibrate user trust and reliance on these technologies. 

In the summer of 2023, Amazon began to test, and eventually incorporated, an AI generated overview of customer reviews associated with each product~\cite{Palmer_2023}. The AI overview opens with ``Customers say,'' followed by a paragraph summary of the customer reviews and ends with a line clarifying that the summary is ``Generated from the text of customer reviews.'' From Amazon's Amazon News outlet, these AI overviews are meant to ``feature key product insights and allow customers to more easily surface reviews that mention certain product attributes'' and are only generated from common themes in multiple verified purchases~\cite{Vaughn_Schermerhorn_2023, Levine_2024}. 
We ground our experiment in Amazon's AI review summaries because e-commerce provides a naturalistic setting in which users must weigh AI-generated summaries against the underlying user-generated information. Purchasing decisions also have direct consequences for users, and Amazon provides an established instance of this increasingly common design pattern. Our experimental stimuli are drawn from real AI overviews and review that users might encounter on the platform.

In e-commerce settings, unlike in physical stores, users do not have physical access to the product before purchase. They are thus dependent on details offered by the merchant, and reviews from previous buyers, to inform their purchasing decisions. A previous study found that 85.57\% of participants stated that they ``read reviews often or very often before they purchase online'' and that a similar percentage of participants compare positive and negative reviews against each other to make their decisions~\cite{lackermair2013importance}. The number of reviews also has a strong effect in engendering user trust of a product before the decision to purchase~\cite{airbnb}. Yet, potential buyers do not consume reviews with complete objectivity; often, customers subconsciously consume reviews as a process of self-verification, wherein reviews are judged more favorably if they align with the user's preexisting biases~\cite{selfverification}. Taken together, these factors--the importance of reviews and the simultaneous subjective nature by which users parse them--pose an interesting question for how AI summaries of the reviews might affect the process by which users evaluate products to purchase. 

\begin{figure*}
    \centering
    \includegraphics[width=\linewidth]{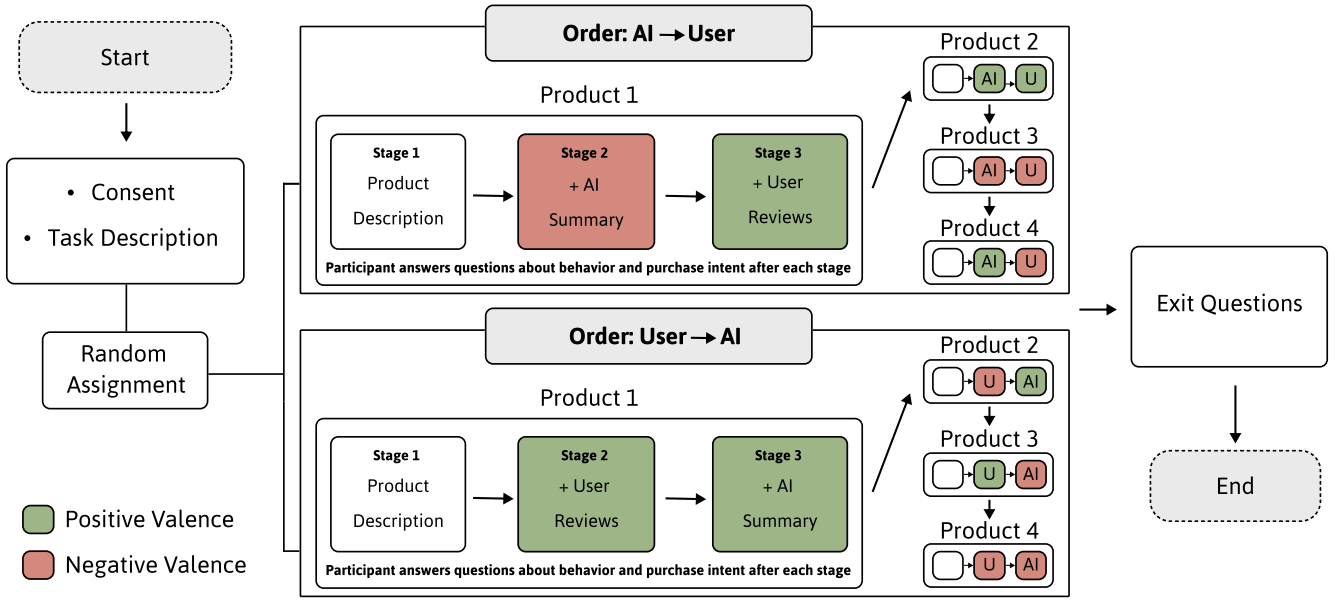}
    \caption{This 2x2x2 experimental setup utilizes a mixed design with three experimental conditions: order of information (AI summary first vs user review first), AI summary valence, and user review valence. Participants are randomly assigned to an order, and asked to provide their perceptions of four products with various information valences in three stages.}
    \Description{Flowchart of the study procedure. After consent and task instructions, participants are randomly assigned to one of two fixed presentation orders: AI summary followed by user reviews, or user reviews followed by AI summary. For each of four products, participants first see the product description, then the first information source, and then the second source, answering attitude and purchase-intent questions after each stage. Across the four products, positive and negative AI summaries and user reviews are varied to cover the different valence combinations. Participants complete exit questions after evaluating all four products.}
    \label{fig:study-overview}
\end{figure*}

\subsection{Study Overview}
In this paper, we conduct a randomized experiment to examine how participants' product attitudes and purchase intent are influenced by AI-generated summaries (AI summaries) and user-generated reviews (user reviews) across positive and negative valences. We present participants with scenarios in which the two sources align and conflict.

We use real product descriptions, user reviews, and AI overviews, introducing them to participants in stages to isolate the influence of each source on participants' attitudes and purchase intent. Our study has two key objectives: to examine how users incorporate AI summaries into their decision making, and to compare the influence of AI summaries with that of the user reviews they summarize. Based on these objectives, we formulate the following research questions. 

We first ask whether AI summaries influence users' judgments: 

\noindent\hangindent=2em\hangafter=1
\textbf{RQ1:} To what extent do AI summaries influence users' attitudes and purchasing intent?\par

Past consumer research has shown that customers compare positive and negative reviews, and that negative reviews can have a more pronounced impact on consumer purchasing decisions than positive reviews~\cite{lackermair2013importance}. Accordingly, we ask:

\noindent\hangindent=2em\hangafter=1
\textbf{RQ2:} To what extent does the valence of an AI summary impact its influence on users?\par

We next compare the influence of AI summaries with that of user reviews, including when the two sources conflict:

\noindent\hangindent=2em\hangafter=1
\textbf{RQ3:} Do participants' judgments align more with AI summaries or with user reviews, and how does this change when the two sources conflict?\par

Finally, because interfaces determine which source users encounter first, we examine how presentation order shapes these effects:

\noindent\hangindent=2em\hangafter=1
\textbf{RQ4:} How does the order in which users encounter AI summaries and user reviews affect their judgment?\par

\subsection{Summary of Contributions}
We present a preregistered, randomized experiment ($N$=278) examining how users integrate AI summaries and user-generated information in an e-commerce setting.
Our analyses reveal a pattern of \emph{conditional reliance}: the influence of AI summaries depends on both their content, particularly their valence, and their placement.
For \textbf{RQ1}, participants generally shifted their judgments in the direction of AI summaries, aligning with the summary's valence.
For \textbf{RQ2}, negative AI summaries were more influential than positive ones.
For \textbf{RQ3}, AI summaries were not more influential than user reviews overall; when the two sources conflicted, participants aligned more strongly with user reviews.
For \textbf{RQ4}, presentation order shaped these effects: AI summaries were more influential when encountered before user reviews, whereas user reviews were similarly influential whether encountered before or after the AI summary.
At the same time, participants consistently rated user reviews as more helpful, even though AI summaries sometimes exerted greater influence on their judgments. This suggests a form of \emph{unrecognized reliance}, in which users may underestimate the influence of AI summaries on their decision making. 
Our findings have implications for how platforms position AI-generated summaries relative to the sources they summarize, both in e-commerce and in AI-mediated information access more broadly.

\section{Related Work}

\subsection{AI-assisted Decision Making}

AI's ability to influence and assist users has previously been examined; users have been found to overrely on AI advice, even in scenarios in which the AI advice is in conflict with their own assessment and expert advice, to the extent of negatively affecting themselves and their cooperative partners~\cite{klingbeil2024trust}, motivating the need for continuing research on how users perceive AI assistance in decision making and reconcile that advice with other sources of information.~\citet{yin2019understanding} found that a model's stated and observed accuracies impacted laypeople's trust when making decisions on outcomes of speed dating events,~\citet{zhang2020effect} found that displaying a model's confidence score can better calibrate a user's trust in the model in the housing estimation context, and~\citet{li2026guided} found that providing explainable AI with guided reflection reduced over-reliance and improved decision accuracy in diabetes prediction. \textit{We extend these findings to a new domain--e-commerce, investigating how the commercial connotations of purchasing decisions affect a user's use of AI assistance in their decisions}. Additionally, the experimental designs of many of these papers, in which a user is asked to give a judgment and is then given additional information before being prompted to give the judgment again, informed our own experimental setup. 

\subsection{Generative AI Summaries in Online Platforms}
Generative AI has transformed the information access (IA) space.~\citet{informationaccess}, however, identify potential pitfalls of Generative IA, and the process of flattening multiple human-generated sources into a singular higher level summary---significantly, that it may exhibit bias in answers, output ungrounded answers, and lack appropriate friction, potentially leading to cognitive laziness.

Findings in the effect of AI generated summaries have validated these concerns. Research into the AI summaries of search results shows evidence that they overly prioritize certain sources over others~\cite{huang2026answer, dai2024neural, Chapekis_Lieb_2025, sharma2024generative}, and that dense retrievers exhibit biases~\cite{dai2024neural, goyal2026masking}. Users who encounter AI summaries are also more likely not to click on subsequent links~\cite{Chapekis_Lieb_2025}. Taken together, these studies suggest a simultaneous unreliability and overreliance on AI summaries. Similarly,~\citet{govers2026narratives} found that, in the context of social media discourse, AI summaries lead users to incorrectly perceive polarized threads as balanced, and that displaying commenter agreement percentages increased conformity to the majority view. The inclusion of AI generated elements in e-commerce settings has already been found to introduce problematic effects;~\citet{kelly2025understanding} studied 10,000 AI-generated product descriptions on eBay, and found that these descriptions exhibited gender bias, promoting exclusionary assumptions and stereotypes. \textit{In addition to deepening the investigation of AI summaries in the e-commerce domain, we also focus on how users synthesize generative IA with traditional user-generated information to make their decisions. 
}
\subsection{Online Reviews and Social Influence}
Previous work has investigated the confluence of features that motivate consumers' purchasing processes. Online word of mouth is integral in e-commerce settings, where there is an information asymmetry between customers and suppliers~\cite{nelson1974advertising}, since customers cannot physically interact with and verify the quality of the product prior to purchase~\cite{dellarocas2003digitization}. Online reviews have been found to sway user decisions~\cite{Guo2020PositiveEBA, Chen2022TheIOA}. Reputation also plays a significant role in consumer's differentiating power. A study on AirBnB found that the number of reviews a user received had a larger effect on trustworthiness than their average rating~\cite{airbnb}, and a study on Amazon book purchases found that higher proportions of helpful votes and spotlight reviews are correlated with an increase in sales~\cite{chen2008all}. Clearly, reviews are not only evaluated based on their textual content; the provenance of the information, and its reception by other users, also factors into a consumer's overall perception of the product. Furthermore, the valence of a review also has an impact on its effect--the percentage of negative reviews have been found to have a stronger effect on new product sales than that of positive reviews~\cite{cui2012effect}, motivating our investigation into the effects of AI summary valence. Existing work has also highlighted the concerns surrounding generative AI in the e-commerce space, finding that frontier LLMs have a sizeable persuasive ability to steer consumers to select sponsored products~\cite{salvi2026commercial}. \textit{Our work extends these findings to examine the reputation of AI summaries, which have a different provenance and may lack the organic, crowd-sourced perception that engenders trust in user-generated reviews. }

\section{Hypotheses}

We formulate the following hypotheses to address our research questions, based on the findings of prior work: 

\textbf{RQ1} focuses on the effect of AI summaries. Following work on the effects of AI assistance on participant decision making~\citep{zhang2020effect, yin2019understanding}, we hypothesize:

\noindent\hangindent=2em\hangafter=1
\textbf{(H1)} Being shown an AI-generated summary will cause participants to shift their attitudes and purchase intent to align with the summary's valence.\par

\textbf{RQ2 }examines the effect of valence on the influence of the summary. We hypothesize that, as with user-generated reviews, negative AI summaries will have a larger impact than their positive counterparts:

\noindent\hangindent=2em\hangafter=1
\textbf{(H2)} Negative AI summaries will produce larger shifts in user attitudes and purchase intent than positive ones.\par

\textbf{RQ3} introduces the element of user reviews, and compares the influence of the user reviews against that of the AI summaries. We hypothesize:

\noindent\hangindent=2em\hangafter=1
\textbf{(H3A)} Users' attitudes and purchase intent will align more with AI summaries than with user reviews.\par

\noindent\hangindent=2em\hangafter=1
\textbf{(H3B)} When AI summaries and user reviews conflict, users will align more with AI summaries than with user reviews.\par

\textbf{RQ4 }examines the effect of order---whether seeing a source first or second will have an effect on the source's influence. We hypothesize effects in line with anchoring effects:

\noindent\hangindent=2em\hangafter=1
\textbf{(H4)} Users' final judgments will align more with whichever source they encountered first, consistent with anchoring effects.\par
\section{Experimental Design}

We designed a 2x2x2 mixed design experiment to investigate when and how users rely on information about a product's description, AI summary, and user reviews to make their purchasing decisions. Our survey included four products. Following prior work distinguishing products between search products, such as floor coverings and hardware, whose quality and utility are evaluated before purchase, and experience products, such as books or bicycles, which must be experienced to be evaluated~\cite{nelson1970information, selfverification, nakayama2010has}, we chose to focus on search products to minimize the participants' personal preferences when assessing product information. We selected four search products from four different Amazon product departments to represent a variety of price ranges and areas of interest. These products were a camera, a dog leash, a changing table, and garbage bags. We conducted five rounds of pilot studies with departmental colleagues, personal contacts, and a paid sample of 10 participants on Prolific, collectively consisting of 29 participants, to iterate on the presentation of elements, as well as the questions asked to assess participant sentiment. This study was preregistered.\footnote{\url{https://osf.io/fs8rt/overview?view_only=bd019973c6ed4bc6a6c026dce663f9a5}} 

\subsection{Study Procedure}

We conducted our survey through the Qualtrics online survey platform. At the beginning of the survey, participants were asked to read the consent form and provide their consent, and then informed that they would be giving opinions on different products and instructed to answer as honestly as possible, with only the information provided. They then moved into the survey, where participants saw each product in three stages: 

\textit{Stage 1:} In the first stage, the participant is given the prompt: \textit{Your friend has asked you to buy a [product] for their home. Based on the below information about the product, please answer the following questions}. After this, they are given the title, image, and description for a product, screenshotted from the product's actual Amazon page.  

\textit{Stage 2}\textit{:} In the second stage, the participant is given additional information about the product, either the AI summary or the user-generated reviews.

\textit{Stage 3:} In the third stage, the participant is given the last piece of information about the product, the AI summary or the user-generated reviews, depending on which of the two they did not see in Stage 2.

At each stage, participants were prompted to answer the same set of questions so that their answers could be compared across stages. These questions and answer ranges were adapted from prior literature exploring customer satisfaction of consumer products~\cite{loureiro2020norms, kyto2019comparison}. To measure \textit{attitude}, the first question asked \textit{How do you feel about this product?}, which participants could answer using a 5-point Likert scale that ranged from \textit{Very positive} to \textit{Very negative}. To measure \textit{purchase intent,} the second question asked \textit{How likely are you to buy this product for your friend?}, which participants could answer on a 5-point Likert scale that ranged from \textit{I am very unlikely to purchase} to \textit{I am very likely to purchase}. These self-reported metrics allowed us to examine the effect of each addition of information on both participant attitude and purchase intent. 

At the end of the Stage 3 of every product, participants were given the opportunity to mark the specific reviews they found helpful and to rank the product description, AI summary, and user reviews by their helpfulness. They were also given the opportunity to provide an open-ended elaboration on whether they would have liked to have any other information about the product when answering the questions. We concluded the survey with a set of exit questions, which asked about participants' prior Amazon usage, as well as their final rankings of each element's helpfulness in making their decision; this final ranking was supplemented by an open-ended text box where participants could elaborate about the aspects of each element that were most or least helpful in their rankings. 

\subsubsection{Justification for Methodological Choices} We elected to frame the purchase decisions as buying for a friend (as opposed to buying for self) to minimize the effect of the participant's own budget and demand for the displayed product on participants' considerations. Prior consumer research defines two dimensions of choosing-for-others purchasing decisions: social focus and ``whether the chooser is primarily considering the recipient’s consumption preferences or balancing them with the chooser’s own preferences for the situation''~\cite{liu2019framework}. \citet{liu2019framework}  classify the\textit{ recipient focus} and \textit{highlighting recipient's preferences} quadrant as everyday favors/pick-ups, which prioritize the recipient's intentions and de-prioritize the chooser's own preferences. Following this, we intentionally kept the details on the relationship between the hypothetical friend and the participant sparse to keep the social focus on the recipient, rather than the participant's relationship with the recipient; as for the second dimension, the task of buying a product for another inherently centers the recipient's preferences (as opposed to selecting a venue for a shared meal). After feedback on Pilots 2 to 3, we removed a question on whether the participant would buy the product for themselves.

\subsection{Participants}

For our preregistered experiment, we recruited participants from the crowdsourcing platform Prolific, and limited participants to those who (1) were at least 18 years old, (2) resided in the United States, and (3) were fluent in English. We performed a power analysis using the G*power software for a medium effect size with 80\% power, using a significance level of 0.05, yielding a sample size of 277 participants. In our recruitment we planned to run up to 300 surveys to account for participants whose responses would be discarded, and ultimately administered 290 surveys out of which 12 were dropped. After this, we had 278 participants for our analyses. After pilot testing (Pilot 5), we determined an estimated completion time of 10 minutes, and excluded participants who took the pilot from also participating in the main study. Participants were compensated \$2.50 USD for their work, using an hourly rate of \$15 USD. The survey itself was hosted on the survey platform Qualtrics. We included two attention checks, in accordance with Prolific's attention and comprehension check policy, and only screened participants out if they failed both attention checks. This study was approved by the Institutional Review Board (IRB) of our institution, and we have included the participant demographics in Appendix A.1. 

\begin{figure}
    \centering
    \includegraphics[width=\linewidth]{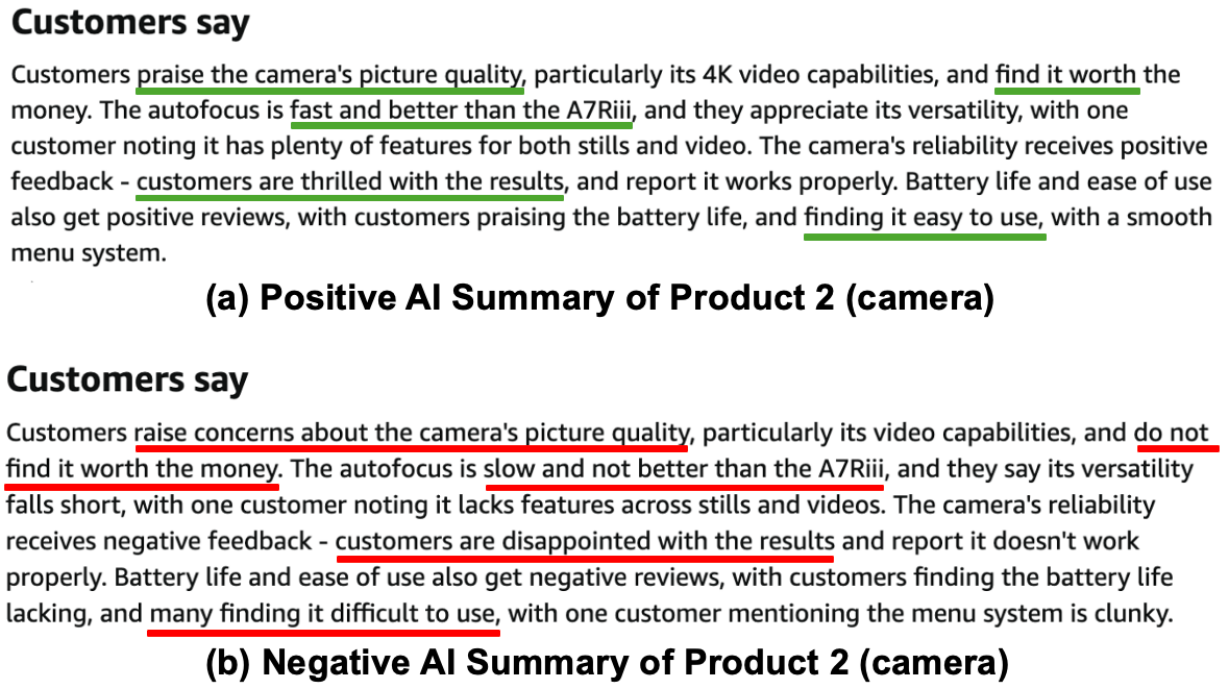}
    \caption{Summaries were perturbed by taking existing summaries and altering the valence of the features, preserving the structure and syntax. The underlines mark a selection of the words and phrases that have been altered.}
    \Description{Two versions of the AI-generated review summary for the camera are shown with nearly identical sentence structure. The positive version praises attributes including picture quality, autofocus, versatility, reliability, battery life, and ease of use, while the negative version expresses concerns about the same attributes. Underlined phrases identify examples of wording whose valence was changed between versions, illustrating that the manipulation preserves the topics and syntax while reversing evaluative language.}
    \label{fig:ai-summary-example}
\end{figure}

\subsection{Experimental Conditions and Operationalization}

Our experimental design manipulated three experimental conditions: the order in which participants encountered information, the valence of the AI summary, and the valence of the user reviews. We now introduce each of these conditions, as well as other features derived from these primary features, which are the variables used in our regression models. Reviews and summaries were collected in April and May of 2025. 

\textbf{Order:} Participants were randomly assigned to see either the AI summary or user reviews during Stage 2, and then to see the other, unseen type of information during Stage 3. This allowed us to account for anchoring effects---whether participants' answers tended to remain aligned with their initial opinions regardless of the information type---and, alternatively, whether participants' reliance and alignment on an information type varied depending on whether they saw it first. We encode this condition as $\textsc{Order}=1$ when the AI summary appeared before the user reviews, and $\textsc{Order}=0$ when the user reviews appeared before the AI summary.

\textbf{AI Summary valence:} For each product, participants could see either a positive or negative AI summary. Rather than generating summaries separately on a pool of positive and negative reviews, which may have focused on different aspects of the product in each summary and introduced potential bias, a product's AI summaries were produced by manually perturbing its existing AI overview. This ensured that the summaries highlighted the same aspects of the product, with the only difference being the judgment on that aspect. Positive or negative judgments were replaced by judgments of the opposite valence--i.e, a product described as being ``hard to set up'' in the negative AI summary instead being described as having ``easy to set up'' in the positive AI summary. A full example of these perturbations can be seen in Figure 2, and the full set of AI summaries can be found in Appendix A.2. We encode this condition as $\textsc{Val}_{AI}=+1$ for positive AI summaries and $\textsc{Val}_{AI}=-1$ for negative AI summaries.

\begin{figure*}
    \centering
    \includegraphics[width=\linewidth]{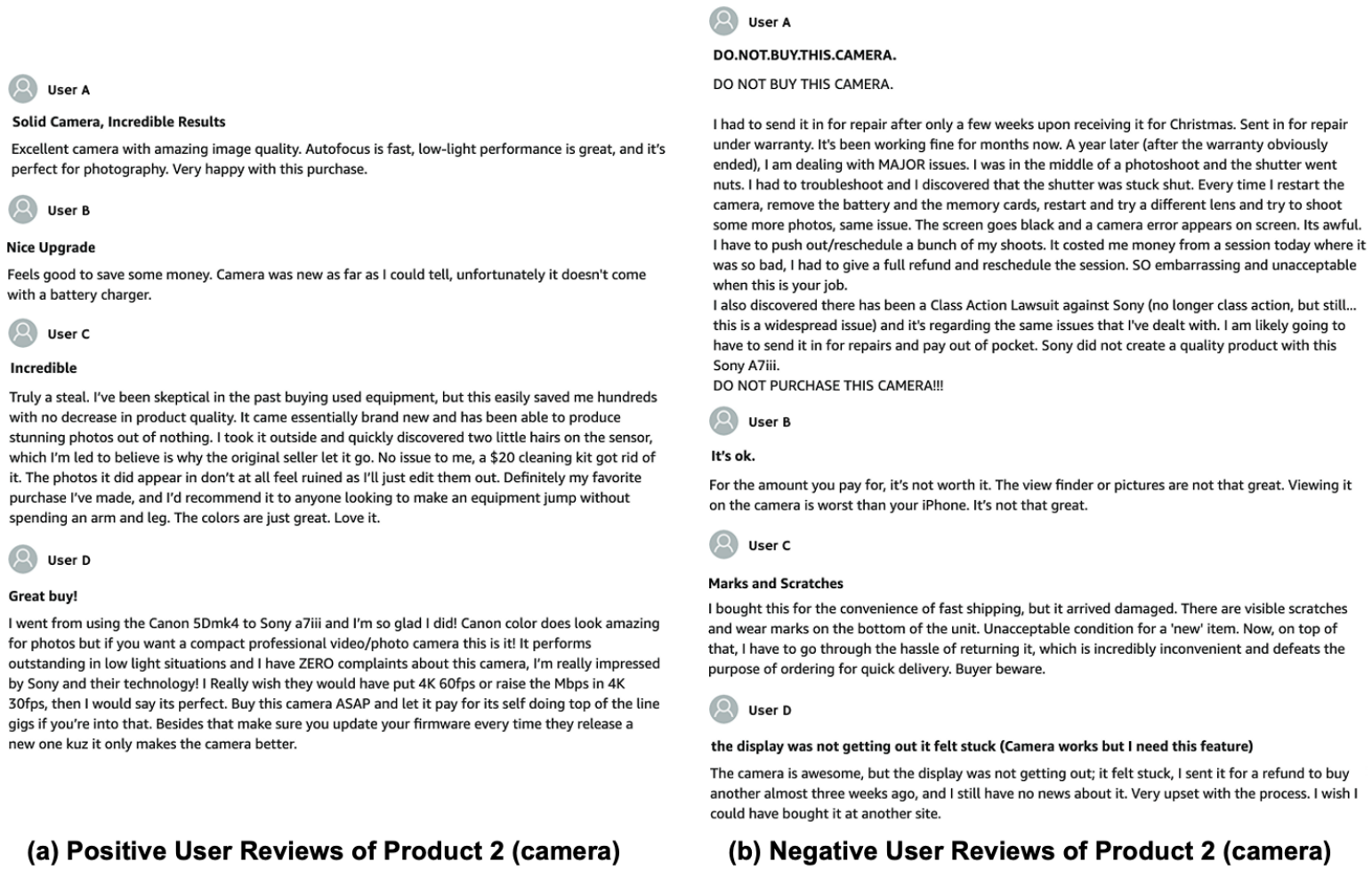}
    \caption{User reviews were collected from the Amazon reviews for each product. A positive set of reviews consisted of two top sorted five-star reviews and top sorted two four-star reviews. Likewise, a negative set of reviews consisted of two top sorted one-star reviews and two top sorted two-stars reviews.}
    \Description{Two sets of four Amazon customer reviews for the camera are displayed side by side. The positive set contains highly rated reviews that generally praise image quality, value, and camera performance. The negative set contains low-rated reviews describing problems such as malfunctioning equipment, poor value, scratches or damage, and display or viewfinder issues. The figure illustrates the user-generated review stimuli used to manipulate review valence.}
    \label{fig:user-reviews-ex}
\end{figure*}

\textbf{User Review valence:} For each product, participants could see either a positive or negative set of user reviews. A positive set of reviews consisted of two five-star reviews and two four-star reviews, taking from the top sorted five-star and four-star reviews using Amazon's ranking algorithm, as these would be the first positive reviews that users would encounter had they been searching in the wild. Likewise, a negative set of reviews consisted of two one-star reviews and two two-stars reviews. Following prior literature distinguishing between the effects of textual reviews and star ratings, we only included the textual component of a given review~\cite{selfverification, airbnb}. An example of a set of user reviews can be seen in~\ref{fig:user-reviews-ex}, and the full collected reviews can be found in Appendix \ref{Appendix-reviews}. We encode this condition as $\textsc{Val}_{user}=+1$ for positive review sets and $\textsc{Val}_{user}=-1$ for negative review sets.

Following pilot testing (Pilot 2) wherein testers expressed confusion about obvious valence bias when given fixed valence combinations, we fixed only the \textsc{Order} for every participant, so that each would either see the AI summary first or the user reviews first for every product. Participants otherwise saw every combination of positive and negative AI summaries and user reviews across the four products. 

\textbf{Conflict:} Because participants saw both an AI summary and user reviews for each product, we also derive a conflict indicator, $\textsc{Conflict}=1$ when $\textsc{Val}_{AI} \neq \textsc{Val}_{user}$, and $\textsc{Conflict}=0$ when the two sources had aligned valences.

\textbf{Influence and Directional Influence:} Our experimental design also allows us to define source-specific influence measures. For each outcome $y \in \{\text{attitude}, \text{purchase intent}\}$, we define the influence of a source $s \in \{\text{AI}, \text{user}\}$ as the change in the participant's response immediately after that source was introduced:
$$\textsc{Infl}_{s} = y_{\text{post}(s)} - y_{\text{pre}(s)}.$$
To compare shifts toward or away from the presented information across positive and negative stimuli, we additionally define directional influence as
$$\textsc{DirInfl}_{s}
=
\textsc{Infl}_{s} \times \textsc{Val}_{s}.$$
Positive values therefore indicate movement in the direction implied by the source, while negative values indicate movement against it. This operationalization allow us to estimate whether participants shift toward AI summaries and user reviews, whether negative information produces larger shifts than positive information, whether conflicting sources differentially affect judgments, and whether first-seen information anchors final responses.

\subsection{Regression Models}

In accordance with our preregistration, we analyze participants' responses with linear mixed-effects models that include crossed random intercepts for participant and product. Each participant evaluated four products, and each product was evaluated in three stages; the random intercepts account for this repeated-measures structure. We therefore include random intercepts for both participant and product in all primary models. We estimate all models separately for the two main outcomes described before: \textbf{product attitude} (measured by responses to \textit{How do you feel about this product?}), and \textbf{purchase intent} (measured by responses to \textit{How likely are you to buy this product for your friend?}). In every model that includes the prior rating $\textsc{Base}$, we mean-center $\textsc{Base}$ within the estimation sample. This does not change slope coefficients---$\textsc{Base}$ is never interacted with other predictors---but makes the intercept the expected directional influence at the average prior rating rather than at the impossible Likert value of 0. Unless noted otherwise, we interpret coefficients after a Bonferroni correction over the predictors within each model, matching the multiplicity adjustment specified in the preregistration. Table~\ref{tab:confirmatory-coefs} reports the full confirmatory estimates; exploratory follow-ups (including the hurdle models in RQ3) are labeled as such and use participant-clustered standard errors.

\begin{table*}[t]
\centering
\sffamily
\caption{Confirmatory mixed-effects estimates for directional influence (crossed random intercepts for participant and product). Stars reflect Bonferroni-adjusted significance within each model (H1: $m{=}1$; H2: $m{=}4$; H3/H4: $m{=}5$); intercepts and baseline use unadjusted $p$. $^{*}p<.05$, $^{**}p<.01$, $^{***}p<.001$.}\Description{Confirmatory mixed-effects regression estimates for directional influence on product attitude and purchase intent. For H1, AI summaries significantly shift both outcomes toward the summary. For H2, negative AI summaries produce greater directional influence than positive summaries, and AI influence is substantially larger when the AI summary is presented first; all corresponding coefficients are significant at p < .001. In the pooled H3/H4 model, the main effect of source is not significant, while conflict increases directional influence toward user reviews relative to AI summaries, and presenting the AI first produces greater overall directional influence. Models include centered baseline ratings and crossed random intercepts for participant and product.}
\label{tab:confirmatory-coefs}
\setlength{\tabcolsep}{4pt}
\renewcommand{\arraystretch}{1.15}
\resizebox{0.8\textwidth}{!}{%
\begin{tabular}{llcccc}
 &  & \multicolumn{2}{c}{\textbf{Attitude}} & \multicolumn{2}{c}{\textbf{Purchase intent}} \\
\cmidrule(lr){3-4}\cmidrule(lr){5-6}
 \rowcolor{blue!10}\textbf{Model} & \textbf{Term} & Est.\ (SE) & $p$ & Est.\ (SE) & $p$ \\
\midrule
 \rowcolor{gray!10}\textbf{H1} (AI only) & Intercept & $0.640^{***}$ (0.033) & $<.001$ & $0.605^{***}$ (0.031) & $<.001$ \\
 & Baseline (centered) & $0.182^{***}$ (0.023) & $<.001$ & $0.166^{***}$ (0.021) & $<.001$ \\
\midrule
 \rowcolor{gray!10}\textbf{H2} (AI only) & Intercept & $0.387^{***}$ (0.044) & $<.001$ & $0.382^{***}$ (0.041) & $<.001$ \\
 & $\textsc{Val}_{\mathrm{AI}}$ & $-0.232^{***}$ (0.033) & $<.001$ & $-0.177^{***}$ (0.032) & $<.001$ \\
&  \graycell{$\textsc{Order}$} & \graycell{$0.503^{***}$ (0.064)} & \graycell{$<.001$} & \graycell{$0.442^{***}$ (0.059)} & \graycell{$<.001$} \\
 & $\textsc{Val}_{\mathrm{AI}}\times\textsc{Order}$ & $-0.362^{***}$ (0.046) & $<.001$ & $-0.349^{***}$ (0.045) & $<.001$ \\
&  \graycell{Baseline (centered)} & \graycell{$0.124^{***}$ (0.021)} & \graycell{$<.001$} & \graycell{$0.120^{***}$ (0.019)} & \graycell{$<.001$} \\
\midrule
 \rowcolor{gray!10}\textbf{H3A/H3B/H4} (pooled) & Intercept & $0.594^{***}$ (0.055) & $<.001$ & $0.587^{***}$ (0.054) & $<.001$ \\
 & $\textsc{Source}$ (AI) & $-0.102$ (0.067) & $0.658$ & $-0.117$ (0.065) & $0.361$ \\
&  \graycell{$\textsc{Conflict}$} & \graycell{$0.672^{***}$ (0.069)} & \graycell{$<.001$} & \graycell{$0.607^{***}$ (0.068)} & \graycell{$<.001$} \\
 & $\textsc{Source}\times\textsc{Conflict}$ & $-0.589^{***}$ (0.095) & $<.001$ & $-0.512^{***}$ (0.092) & $<.001$ \\
 &  \graycell{$\textsc{Order}$} & \graycell{$0.250^{***}$ (0.050)} & \graycell{$<.001$} & \graycell{$0.221^{***}$ (0.050)} & \graycell{$<.001$} \\
 & Baseline (centered) & $0.176^{***}$ (0.026) & $<.001$ & $0.201^{***}$ (0.023) & $<.001$ \\
\bottomrule
\end{tabular}%
}
\end{table*}

\section{RQ1: Influence of AI Summaries}

To test \textbf{H1}, that being shown an AI-generated summary causes users to change their answers to align with the summary, we isolated the responses participants gave after seeing the AI summaries. Following the preregistered specification, we model directional influence of the AI summary as
\begin{align}
\label{eq:h1}
\textsc{DirInfl}_{ij,\text{AI}} = \beta_0 + \beta_1 \textsc{Base}^{\prime}_{ij,\text{AI}} + u_i + v_j + \epsilon_{ij},
\end{align}
where $i$ indexes participants, $j$ indexes products, $u_i$ and $v_j$ are participant- and product-level random intercepts, and $\textsc{Base}^{\prime}_{ij,\text{AI}} = \textsc{Base}_{ij,\text{AI}} - \overline{\textsc{Base}}$ is the mean-centered prior rating (the response immediately before the AI summary). Equivalently, if one starts from an uncentered covariate $\textsc{Base}$,
\begin{align}
\beta_0 + \beta_1 \textsc{Base}
= \bigl(\beta_0 + \beta_1\overline{\textsc{Base}}\bigr) + \beta_1 \textsc{Base}^{\prime},
\end{align}
so mean-centering reparameterizes the intercept and $\beta_0$ is the expected directional shift toward the AI summary at the average prior rating---the natural confirmatory test of \textbf{H1}.

Upon fitting the model, we find that \textbf{H1 is supported.} The intercept is positive and significant both for attitude ($\beta_0 = 0.640$, SE $= 0.033$, $p < .001$) and purchase intent ($\beta_0 = 0.605$, SE $= 0.031$, $p < .001$) (See Table~\ref{tab:confirmatory-coefs}). Participants shift their ratings in the direction implied by the AI summary. While not the confirmatory test of \textbf{H1}, we additionally find that prior ratings predict larger subsequent shifts toward the summary (attitude: $\beta_1 = 0.182$, SE $= 0.023$, $p < .001$; purchase intent: $\beta_1 = 0.166$, SE $= 0.021$, $p < .001$). 

\section{RQ2: Overall Valence of AI Summaries}

To test \textbf{H2}, that users' answers align more closely with negative AI summaries than positive AI summaries, we model the directional influence of AI as:
\begin{align}
    &\textsc{DirInfl}_{ij,\text{AI}} = \beta_0 + \beta_1 \textsc{Val}_{ij,\text{AI}} + \beta_2 \textsc{Order}_{i}\nonumber\\ &+ \beta_3 \left(\textsc{Val}_{ij,\text{AI}} \times \textsc{Order}_{i}\right) + \beta_4 \textsc{Base}^{\prime}_{ij,\text{AI}} + u_i + v_j + \epsilon_{ij} 
\end{align}
where $\textsc{Base}^{\prime}$ is mean-centered as in RQ1, and coefficient $\beta_1$ captures whether directional alignment differs between positive and negative AI summaries. As with RQ1, we similarly isolated the responses participants gave after seeing AI summaries.

After fitting the models, we found that the model indicates a significant effect of AI summary valence for both purchase intent ($\beta_1$ = $-$ 0.177, SE = 0.032, p $<$ .001) and attitude directional influence ($\beta_1$ = $-$ 0.232, SE = 0.033, p $<$ .001). Since $\textsc{Val}_{\text{AI}}=+1$ for positive AI summaries and $-1$ for negative AI summaries, a significantly negative $\beta_1$ indicates that negative AI summaries produce greater directional alignment than positive AI summaries. This result affirms that participants are more affected by negative summaries than positive ones. 
\textbf{H2 is supported}.

Crucially, the interaction term tests whether this valence effect depends on whether the AI summary was shown before or after the user reviews. The model reveals a significant interaction between AI summary valence and presentation order for both purchase intent ($\beta_3$ = $-$0.349, SE = 0.045, p $<$ .001) and attitude ($\beta_3$ = $-$0.362, SE = 0.046, p $<$ .001) directional influence. These results suggest that negative AI summaries do produce greater directional alignment than positive AI summaries, and that this effect is dependent on whether the AI summary was shown before or after the user reviews. The effect of AI valence is significantly stronger when AI summaries are shown before user reviews, rather than when they are shown after user reviews.

\begin{table*}[t]
\centering\sffamily
\caption{Summary of hypotheses and outcomes.}\Description{Summary of the five preregistered hypotheses and their outcomes. H1, predicting that AI summaries shift attitudes and purchase intent toward their valence, is supported. H2, predicting stronger effects for negative than positive AI summaries, is supported. H3A, predicting greater overall alignment with AI summaries than user reviews, is not supported. H3B, predicting greater alignment with AI when the sources conflict, is reversed: participants instead align more with user reviews. H4, predicting greater alignment with the source encountered first, is supported.}
\label{tab:hypotheses}
\setlength{\tabcolsep}{6pt}
\renewcommand{\arraystretch}{1.3}
\begin{tabular}{p{11cm}l}
 \rowcolor{blue!10}\textbf{Hypothesis}& \textbf{Outcome} \\
\midrule
 \rowcolor{gray!10}\textbf{H1}: Being shown an AI-generated summary will cause participants to shift their attitudes and purchase intent to align with the summary’s valence& Supported \\[4pt]
\textbf{H2}: Negative AI summaries will produce larger shifts in
user attitudes and purchase intent than positive ones.& Supported\\[4pt]
  \rowcolor{gray!10}\textbf{H3A}: Users’ attitudes and purchase intent will align more with AI summaries than with user reviews.& Not Supported\\[4pt]
 \textbf{H3B}: When AI summaries and user reviews conflict, users will align more with AI summaries than with user reviews.& Reversed \\[4pt]
  \rowcolor{gray!10}\textbf{H4}: Users’ final judgments will align more with whichever source they encountered first, consistent with anchoring
effects.& Supported \\
\bottomrule
\end{tabular}
\end{table*}

\section{RQ3: Alignment with User Reviews}

To test \textbf{H3A} and \textbf{H3B}, we take each participant-product pair and consider two observations: the directional influence of the AI summary and  the directional influence of the user reviews, and define $\textsc{Source}_{ij,s}=1$ for AI summaries and $\textsc{Source}_{ij,s}=0$ for user reviews. Then, we estimate:
\begin{align}
\label{eq:h3h4}
    &\textsc{DirInfl}_{ij,s} = \beta_0 + \beta_1 \textsc{Source}_{ij,s} + \beta_2 \textsc{Conflict}_{ij} \nonumber\\&+ \beta_3 \left(\textsc{Source}_{ij,s} \times \textsc{Conflict}_{ij}\right) + \beta_4 \textsc{Order}_{i} \nonumber\\&+ \beta_5 \textsc{Base}^{\prime}_{ij,s} + u_i + v_j + \epsilon_{ij,s}
\end{align}
where $\textsc{Base}^{\prime}$ is again mean-centered within the estimation sample, and $\beta_1$ tests \textbf{H3A}, whether participants align more strongly with AI summaries than with user reviews overall. The interaction coefficient $\beta_3$ tests \textbf{H3B}: whether the relative influence of AI summaries compared to user reviews is larger when the two sources conflict. 

Upon fitting the models, we found that conflict, order, previous answer, and the interaction of source and conflict to have significant effects for both purchase intent and attitude directional influence. As neither $\beta_1$ were found to have significant effects, we conclude that source alone does not have an effect on directional influence. \textbf{H3A was thus not supported}. 
The interaction between source and conflict was significant for both the attitude ($\beta_3$ = $-$ 0.589, SE = 0.095, p $<$ .001) and purchase intent models ($\beta_3$ = $-$ 0.512, SE = 0.092, p $<$ .001). Since a significantly negative $\beta_3$ indicates that conflict increases a participant's relative alignment with user reviews compared to AI summaries, we find that \textbf{H3B is not supported}, and in fact the \textit{opposite of H3B appears to be true}. This indicates that the influence of AI summaries and user reviews are contingent on whether they conflict with the information already given---and that conflict actually increases a participant's relative alignment with user reviews compared to AI summaries. 

\begin{table}[t]
\centering
\sffamily
\caption{Percentage of observations showing no change ($\textsc{DirInfl} = 0$) by measured source and whether the AI overview and user reviews were \textit{aligned} or \textit{conflicting}.}\Description{Percentage of observations in which participants did not change their rating after seeing an information source. For attitude, no-change rates are 43.2 percent for user reviews and 55.0 percent for AI summaries when the sources align, versus 22.5 percent and 46.2 percent, respectively, when they conflict. For purchase intent, the corresponding rates are 44.6 and 59.7 percent when aligned and 24.1 and 48.4 percent when conflicting. Conflict therefore makes participants substantially more likely to revise their ratings in response to user reviews, while the reduction is smaller for AI summaries.}
\label{tab:no-change}
\setlength{\tabcolsep}{6pt}
\renewcommand{\arraystretch}{1.2}
\begin{tabular}{llcc}
 & & \multicolumn{2}{c}{\textbf{Conflict}} \\
\cmidrule(lr){3-4}
 \rowcolor{blue!10}\textbf{Outcome} & \textbf{Source} & Aligned & Conflict \\
\midrule
\multirow{2}{*}{Attitude} 
 & \graycell{User reviews} & \graycell{43.2} & \graycell{22.5} \\
 & AI overview  & 55.0 & 46.2 \\
\midrule
\multirow{2}{*}{Purchase intent}
 & \graycell{User reviews} & \graycell{44.6} & \graycell{24.1} \\
 & AI overview  & 59.7 & 48.4 \\
\bottomrule
\end{tabular}
\end{table}

\subsection{Deeper Analysis with Two-Part Hurdle Models}
Upon further investigation, we observe that the directional-influence outcome variable shows a sharp concentration at zero. Depending on the condition, between 23\% and 60\% of observations show no change at all (see Table~\ref{tab:no-change}), with the directional influence of the AI summaries where the sources agreeing showing the highest zero rate ($\sim$60\%) while the directional influence of the user reviews where the sources conflict showing the lowest zero rate ($\sim$23\%). Therefore, we find that a single linear model fit to this outcome conflates two distinct decisions: (i) whether a participant revises their evaluation at all, and (ii) how far they revise given that they do, each of which may respond differently to source and conflict. To unpack the earlier result from the preregistered model, we additionally estimate a two-part hurdle model~\cite{cragg1971some,mullahy1986specification} that separates these two.

\begin{table*}[t]
\centering
\small
\sffamily
\caption{Exploratory two-part hurdle models for RQ3 (participant-clustered SEs). Initiation is a logit for whether the rating changed; magnitude is OLS of directional influence among movers. For initiation, columns report $p$ and the odds ratio (OR). $^{*}p<.05$, $^{**}p<.01$, $^{***}p<.001$.}
\Description{Exploratory two-part hurdle models decomposing directional influence into whether participants changed their rating at all and, among those who changed, the magnitude of the shift. In the initiation models, AI summaries are less likely than user reviews to trigger a rating change, while conflict substantially increases the odds of changing, especially in response to user reviews. The negative source-by-conflict interaction indicates that conflict increases revision less for AI summaries than for user reviews. Seeing the AI summary first and having a higher baseline rating are also associated with greater odds of revision. In the magnitude models, there is no significant main effect of information source, but conflict increases shifts toward user reviews; the negative source-by-conflict interaction shows that this increase is substantially weaker for AI summaries. Presentation order and baseline ratings also positively predict shift magnitude. These patterns are consistent across attitude and purchase intent.}
\label{tab:hurdle-coefs}
\setlength{\tabcolsep}{4pt}
\renewcommand{\arraystretch}{1.15}
\resizebox{0.8\textwidth}{!}{%
\begin{tabular}{llcccc}
 &  & \multicolumn{2}{c}{\textbf{Attitude}} & \multicolumn{2}{c}{\textbf{Purchase intent}} \\
\cmidrule(lr){3-4}\cmidrule(lr){5-6}
\rowcolor{blue!10} \textbf{Part} & \textbf{Term} & Est.\ (SE) & $p$ / OR & Est.\ (SE) & $p$ / OR \\
\midrule
\rowcolor{gray!10}Initiation (logit) & Intercept & 0.025 (0.096) & 0.797 / 1.03 & -0.030 (0.094) & 0.751 / 0.97 \\
 & $\textsc{Source}$ (AI) & -0.462$^{***}$ (0.136) & $<.001$ / 0.63 & -0.583$^{***}$ (0.133) & $<.001$ / 0.56 \\
 & \graycell{$\textsc{Conflict}$} & \graycell{0.974$^{***}$ (0.128)} & \graycell{$<.001$ / 2.65} & \graycell{0.948$^{***}$ (0.123)} & \graycell{$<.001$ / 2.58} \\
 & $\textsc{Source}\times\textsc{Conflict}$ & -0.616$^{***}$ (0.180) & $<.001$ / 0.54 & -0.477$^{**}$ (0.166) & 0.004 / 0.62 \\
 & \graycell{$\textsc{Order}$} & \graycell{0.483$^{***}$ (0.095)} & \graycell{$<.001$ / 1.62} & \graycell{0.452$^{***}$ (0.097)} & \graycell{$<.001$ / 1.57} \\
 & Baseline & 0.101$^{**}$ (0.038) & 0.008 / 1.11 & 0.193$^{***}$ (0.039) & $<.001$ / 1.21 \\
\midrule
\rowcolor{gray!10}Magnitude (OLS) & Intercept & 1.203$^{***}$ (0.074) & $<.001$ & 1.200$^{***}$ (0.078) & $<.001$ \\
 & $\textsc{Source}$ (AI) & -0.050 (0.090) & 0.577 & 0.039 (0.090) & 0.665 \\
 & \graycell{$\textsc{Conflict}$} & \graycell{0.481$^{***}$ (0.089)} & \graycell{$<.001$} & \graycell{0.435$^{***}$ (0.088)} & \graycell{$<.001$} \\
 & $\textsc{Source}\times\textsc{Conflict}$ & -0.511$^{***}$ (0.113) & $<.001$ & -0.550$^{***}$ (0.112) & $<.001$ \\
 & \graycell{$\textsc{Order}$} & \graycell{0.248$^{***}$ (0.073)} & \graycell{$<.001$} & \graycell{0.224$^{**}$ (0.075)} & \graycell{0.003} \\
 & Baseline & 0.309$^{***}$ (0.033) & $<.001$ & 0.311$^{***}$ (0.034) & $<.001$ \\
\bottomrule
\end{tabular}%
}
\end{table*}

The first part models \emph{initiation}, a logistic regression of whether the rating changed in either direction at all,
\begin{align}
\label{eq:hurdle-init}
&\log \frac{\Pr(\textsc{Moved}_{ij,s}=1)}{1-\Pr(\textsc{Moved}_{ij,s}=1)} = \alpha_0 + \alpha_1 \textsc{Source}_{ij,s} \nonumber\\&+ \alpha_2 \textsc{Conflict}_{ij} + \alpha_3 (\textsc{Source}\times\textsc{Conflict})_{ij,s} \nonumber\\&+ \alpha_4 \textsc{Order}_i + \alpha_5 \textsc{Base}^{\prime}_{ij,s}
\end{align}
where $\textsc{Moved}_{ij,s} = \mathbf{1}[\textsc{DirInfl}_{ij,s} \neq 0]$. Each coefficient $\alpha$ is a log-odds, and $\exp(\alpha)$ are the corresponding odds ratio reported below. The second part models \emph{magnitude}, an OLS regression of signed directional influence among movers only ($\textsc{Moved}=1$), with the same predictors but separately estimated coefficients $\delta_0,\dots,\delta_5$. Because the outcome is signed, the directional interpretation---movement \emph{toward} the source---is carried by this second part. We use participant-clustered standard errors for both parts.

The decomposition shows that conflict amplifies the influence of user reviews on \emph{both} outcomes, while the effect of AI summaries remains low as before. From the first part (initiation), we find that AI summaries are less likely than user reviews to provoke any revision when sources agree (purchase: $\alpha_1$=--0.583, $\mathrm{OR}$=0.56, $p<$.001; attitude: $\alpha_1$= --0.462, $\mathrm{OR}$ = 0.63, $p<$.001). At the same time, presence of conflict in the valence of sources raises the odds of revising in response to user reviews by roughly $2.6\times$ (purchase: $\alpha_2$= 0.948, $\mathrm{OR}$ = 2.58, $p<$.001; attitude: $\alpha_2 = 0.974$, $\mathrm{OR} = 2.65$, $p<.001$). However, this effect is not as pronounced for AI summaries (purchase: $\alpha_3$=--0.477, $p =$.004; attitude: $\alpha_3$=--0.616, $p=$.001). From the second part (magnitude), we find that among participants who did revise their ratings, there is no baseline difference between sources when they agree (purchase: $\delta_1 = 0.039$, $p = .665$; attitude: $\delta_1 = -0.050$, $p = .577$), but conflict again increases the magnitude of shifts toward user reviews (purchase: $\delta_2 = 0.435$, $p<.001$; attitude: $\delta_2 = 0.481$, $p<.001$) far more than toward AI summaries (purchase: $\delta_3 = -0.550$, $p<.001$; attitude: $\delta_3 = -0.511$, $p<.001$).

\begin{figure}[t]
    \centering
    \includegraphics[width=\linewidth]{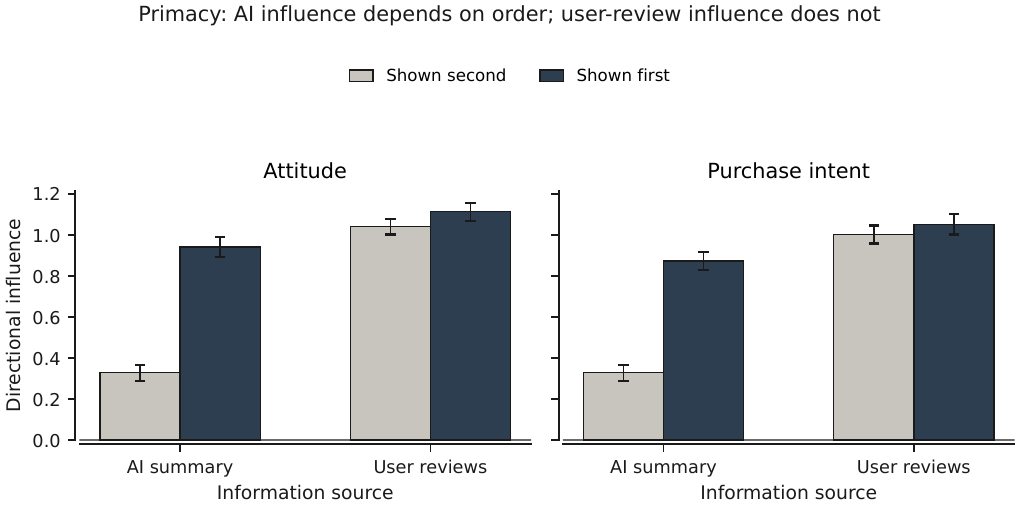}
    \caption{Mean directional influence by source and whether that source was shown first or second. Error bars are participant-clustered standard errors. AI influence depends strongly on order; user-review influence does not.}
    \Description{Two grouped bar charts show mean directional influence for product attitude and purchase intent. For both outcomes, AI summaries have substantially greater influence when shown first than when shown second. In contrast, user reviews have similarly high influence whether shown first or second. Participant-clustered standard-error bars are small relative to the difference between the two AI-summary conditions. The pattern shows a strong primacy effect for AI summaries but little ordering effect for user reviews.}
    \label{fig:order-source}
\end{figure}

\section{RQ4: Ordering Effects}

Finally, to test \textbf{H4}, we use the coefficient $\beta_4$ in equation \eqref{eq:h3h4}. Since $\textsc{Order}=1$ when the AI summary is shown before the user reviews and $\textsc{Order}=0$ when the user reviews are shown before the AI summary, the coefficient on $\textsc{Order}$ captures whether participants show greater directional alignment when the AI summary is encountered first. We find that order has a significant effect for both the attitude ($\beta_4$ = 0.250, SE = 0.050, p $<$ .001) and purchase intent model ($\beta_4$ = 0.221, SE = 0.049, p $<$ .001). The positive coefficients indicate that participants exhibited greater directional alignment when the AI summary was encountered first. \textbf{H4 is supported. }

As an exploratory follow-up, we decompose this order effect by source (Figure~\ref{fig:order-source}). The impact of order is much greater for AI summaries than user reviews: AI summaries have more substantial effect on ratings when shown before the user reviews as opposed to when they are shown after. In contrast, user reviews influence users equally, and are not contingent on whether they are shown before or after the AI summary. This suggests that AI summaries can anchor users' opinions if they are presented before user generated information, but that they can be swayed after user generated information is presented.

\section{Stated vs Revealed Reliance}

We also conducted an exploratory analysis of users' stated preferences of the AI summary, user reviews, and product descriptions. This analysis was not specified in advance, 
so we treat it as exploratory. At the end of each product, users were prompted to rank the helpfulness of the three toward making their decisions. In this section, we compare whether participants' stated preferences aligns with their actual reliance on the given information sources. 

\subsection{Stated preferences}
To begin with, we sought to establish whether participants expressed any significant preference at all for any information source. A Friedman test revealed a significant difference in participants' rankings of the three information sources, $\chi^2(2) = 729.12$, $p < .001$. Post hoc pairwise comparisons with Bonferroni correction showed that user reviews were ranked significantly higher than both AI summaries ($p < .001$) and product descriptions ($p < .001$). Meanwhile, there was no significant difference between AI summaries and product descriptions ($p = .24$). Mean ranks indicated that user reviews ($M = 1.33$) were consistently ranked as more helpful than product descriptions ($M = 2.31$) and AI summaries ($M = 2.36$), with lower mean ranks indicating greater perceived importance. In terms of elicited preference, participants preferred user generated information to other sources. 

\subsection{Local calibration}
We then moved into probing whether these preferences were correlated with how participants were influenced by the information sources. To assess whether participants' stated preferences aligned with their observed behavior, we computed Spearman correlations between source rankings, with lower values indicating greater perceived importance. We measured influence separately for AI summaries and user reviews. Across both the purchase and attitudinal samples, better rankings were associated with greater observed influence for AI summaries (purchase: $\rho=-0.233$, $p<.001$; attitudinal: $\rho=-0.246$, $p<.001$) and user reviews (purchase: $\rho=-0.220$, $p<.001$; attitudinal: $\rho=-0.221$, $p<.001$). Within a single product, participants could tell which source had moved them, and did so about equally well for AI summaries and user reviews, though the correlations were modest.

\subsection{Calibration gap}
Lastly, we looked at whether there was any \textit{misalignment} between participants' perception of the helpfulness of the information sources and the sources' actual influence upon their decisions by comparing their rankings to the actual directional influence. We found that participants' judgment aligned with the more influential source in 67.4\% of cases for purchase intent judgments and 66.2\% of attitude judgments. However, the remaining disagreements were more asymmetric. Participants rarely over-attributed influence to AI summaries (4.8\% and 5.5\% of observations), but much more frequently attributed greater influence to user reviews even when AI summaries produced more directional influence (27.8\% and 28.3\%). Consistent with this pattern, participants ranked AI summaries above user reviews in only 15.6\% of observations, whereas AI summaries actually exerted the larger influence in approximately 27–28\% of observations. These metrics suggest that participants underestimate the ability of AI summaries to influence their perceptions, especially compared to user reviews. 
Prior work on AI-assisted decision making has largely focused on over-reliance~\cite{selfverification, li2025text, klingbeil2024trust, castelo2019task}, where users defer to AI more than its accuracy warrants. Our finding points to a different concern: participants were influenced by information they said they discounted, without recognizing it.

\subsection{Felt sufficiency and distrust}
After answering the purchase intent and attitude questions for each product, participants were prompted to answer an optional open ended question \textit{Did you feel you had all the information necessary to make the decision?}. The authors then performed a manual review of these open ended statements, doing a binary annotation of whether or not participants found the provided information sufficient. Because these procedures
were not specified in advance, we treat them as qualitative support for the quantitative results rather than as independent confirmation. 

The percentage of participants reporting sufficient information fell from 89.3\% when the two sources agreed to 80.6\% when they conflicted. A clustered logistic regression revealed that conflicting information significantly reduced the likelihood that an item was manually labeled as sufficient ($\beta = -0.70$, SE = 0.19, $p < .001$, OR = 0.50). We then examined whether specific valence configurations differed from the baseline condition (AI summaries and user reviews both negative). Compared with this baseline, the configuration with a negative AI summary followed by positive user reviews significantly reduced sufficiency judgments ($\beta = -1.00$, SE = 0.32, $p = .002$, OR = 0.37), as did configurations with positive user reviews followed by a negative AI summary ($\beta = -0.92$, SE = 0.36, $p = .010$, OR = 0.40). No other configurations significantly differed from the baseline. When the user reviews and AI summaries align, participants were less likely to desire additional information; when they conflicted, the provided information was more often found to be insufficient. 

This is corroborated by the additional information participants identified they would have liked to have---participants who were shown positive reviews expressed a desire to see more negative reviews, and vice versa. Participants also expressed the desire to see rating distributions; seeing as rating distributions are another way, besides AI summaries, of providing an overview of user opinions, further work could explore the difference in perceptions of these two aggregates. Another recurring theme was \textit{expertise,} or the lack of it. One participant, for instance, wrote that ``Since I do not know much about cameras, I inherently do not feel like I have enough information despite the AI overview and the reviews looking very positive,'' and other participants expressed doubt over whether the reviewers had the expertise for their opinions to be trusted. This suggests a possible reason for why users do not prefer the AI summaries---it is harder to verify the expertise of an AI. In situations where the AI summary and user reviews misaligned, corroborating the earlier finding about ranked preferences favoring user reviews, participants expressed their greater trust in user reviews. Participants wrote that ``Actual human feedback is important when making big purchases'' and ``I listen to people over AI.''

\section{Discussion}

\subsection{Placement as a Design Lever}

Our findings on the impact of order on AI summary perception have direct implications for interface design. Both AI summaries and user reviews influenced participants when shown first, while only user reviews could override the influence of the AI summary when shown next. On the other hand, AI summaries shown after the user-generated reviews did little to influence users, indicating that AI summaries are most potent as an initial frame, while user reviews retain corrective power even after an initial judgment is made. Moreover, when user reviews conflict with AI summaries, participants updated toward the user-generated reviews. These observations underscore the importance of information placement within an interface: the order in which a platform presents its information determines the source upon which the user anchors their decision. 

Often times, such as with the current implementation of Google's search overview or the Amazon's product overview, the AI summary is presented before the user-generated results, an arrangement that allows the AI summary to set the initial frame. This design amplifies early anchoring on AI summaries, which is problematized by the previously established trend of users encountering AI summaries and simply not clicking on subsequent results at all~\cite{Chapekis_Lieb_2025}. Even if users update towards user-generated information when it conflicts with AI summaries, they may never get the chance to do so if they take AI summaries at face value and do not seek the user-generated information. Within current interface designs, convenience incentivizes users to anchor onto the AI summaries, which can be biased and unrepresentative~\cite{huang2026answer, goyal2026masking, kelly2025understanding, sharma2024generative, dai2024neural} without engaging with the user-generated information. And yet, the open ended responses suggested that participants preferred user-generated information to AI-generated information---a point of friction to inform future designs.  

Our study has implications for how platforms can balance convenience, anchoring effects, and users' preferences in information access interfaces. Such designs could preserve the convenience of the aggregate summary, while leveraging mechanisms that can help users move toward the aggregate's underlying evidence when closer scrutiny is warranted. They might experiment with surfacing the capabilities and limitations of the AI summary for the user, and including incentives for the user to parse the user-generated information. Platforms might explicitly indicate the limitations of AI summaries rather than opting for a more seamless design, or even present the AI summary after a number of user-generated sources. Such a seamful design might even aim to surface contradicting information sources, producing useful friction that incentivizes users to explore user reviews even after reading the AI summary. Finally, future work could explore building tools to indicate when a summary's valence departs from the distribution of underlying rankings, or flag when the summary and reviews disagree. 

\subsection{Conditional Reliance in Decision-Making}
Relative to prior work on overreliance on AI advice
\cite{klingbeil2024trust,yin2019understanding,zhang2020effect}, our results
suggest that users' reliance on AI in decision making varies based on the conditions under which they were presented. AI summaries matter
most when they are presented first and when their valence is negative; they have much more measured effects when user-generated reviews have already anchored the user's opinion, or when the two sources conflict. Our hurdle analysis shows that introduction of the AI summary after the user reviews often does not initiate any revision in user opinion at all. When the two are juxtaposed together, users' skepticism of AI summaries becomes more prominent. This suggests that users exhibit a \textit{conditional reliance} on AI: rather than a blanket overreliance on AI, users demonstrate more agreement with AI summaries in specific conditions---when they came first, when they were negative---and when they did not have user reviews to contextualize them.

The heightened influence of the negative AI summaries compared to their positive counterparts corroborates existing literature on the outsized effect of negative reviews. Prior work shows generative models have persuasive power over users in e-commerce settings~\cite{salvi2026commercial}, and the conditional reliance on AI summaries suggests that these effects extend to AI summaries of those reviews as well. Platforms, especially commerce related ones, may be incentivized to mitigate the negativity of AI summaries so as not to discourage the user from paying for the product. The generative model synthesizing the AI summaries may be biased to systematically select certain types of information over others, effectively hiding the latter, leading the end user to take away different components than they would have with the user reviews. In such cases, the mitigation or selection bias may not be detectable by the user, causing them to make their purchasing decisions without critical pieces of information, reducing user autonomy. 

\subsection{Unrecognized Reliance on AI Summaries}

Participants largely expressed skepticism over the AI summaries---in stated preference, they ranked it beneath user reviews, and sided with the reviews in cases of conflict. However, participants often were influenced by AI summaries even when they marked them as unhelpful. AI summaries were the more influential source about twice as often as they were ranked as such, suggesting an underestimation of the summaries' influence by the user. Stated preferences for user-generated information do not necessarily prevent AI from shaping decisions; a possible future intervention would be in alerting users to this tendency. 

Existing transparency features, such as confidence scores and AI disclosures, target lowering over-trust of AI. However, in the case of this study, participants did not perceive themselves as particularly trustful of the AI, but were influenced by it nonetheless. Interventions designed to raise skepticism of AI may not be effective in such instances, because the issue is not in misperception of AI---the skepticism is already there---but instead in how much users realize it has already shaped their judgment. Future interventions may need to target awareness of influence rather than level of trust, for instance by showing users, after the
fact, how their stated preference compared to the source that actually moved
them, or perhaps providing these capabilities proactively. Prior work has explored the potential of cognitive forcing functions in making AI-assisted decision more interactive~\cite{buccinca2021trust}; similar interventions with AI summaries, intended to elicit analytical thinking, might aim to prompt the user to reflect on the impact of AI in their decision.

\subsection{Implications for E-Commerce}
In prior work, \citet{lackermair2013importance} posit two phases in customers' search for products: an early phase, in which they are trying to narrow down the potential products, and a later phase, where they ultimately purchase a single product from the narrowed-down selection. Specifically, they argue that ``reading a lot of user reviews for several products is not very efficient in this early phase,'' where ``[c]ustomers need compact and concise information about the products.'' In many ways, AI summaries fulfill this role. Our study addresses what happens as users transition into the second phase, when they begin reading individual reviews to make a final product selection while still having access to the AI summary. 

In this study, we presented participants with four reviews at a time, but a natural follow up would investigate how perceptions and use of AI summaries change when the volume of reviews becomes too large to skim, increasing the value of the summary's convenience. Such a study could focus on the transition between the two phases: when users move from higher-level overviews to more granular information, and how their perceptions of AI impact the timing of that transition. Another mechanism to explore is a deeper integration between AI summaries and user reviews. For instance, summaries could directly link individual claims to the reviews that support them. Such a design, which already appears in systems such as Google's AI overview, might preserve the convenience of an AI summary while better supporting users' preference for user-generated information.

Further, our findings also raise questions about accountability for how products and reviewers are represented through AI summaries. 
Given the stronger influence of negative summaries, disproportionate emphasis on particular complaints could affect whether a product remains under consideration, especially when users do not subsequently examine individual reviews. 
Reviewers' experiences may similarly be condensed into an overall characterization that obscures important context or qualifications in their accounts.
This motivates future work examining the consequences of such representations and developing mechanisms through which misleading or incomplete summaries can be identified and challenged. 

Platforms could, for example, allow users, reviewers, and sellers to flag claims that omit relevant context or misrepresent the underlying reviews, while making the supporting evidence available for inspection. 
Such challenges should be assessed against the underlying reviews, with safeguards to prevent legitimate criticism from being suppressed/removed in response to commercial pressure. 
Further, summaries should also be reassessed as new reviews change the evidence on which they are based.

\subsection{Limitations and Future Directions}
In our experimental design, we elected to construct positive and negative
conditions for both user reviews and AI overviews. However, users on e-commerce
platforms are not limited to four reviews, and can toggle between negative and
positive reviews as they desire. This more limited control setup allowed us to present AI
and user-generated information as directly contradicting each other, thus
providing insight into how participants reconcile the two to make their
decisions. However, participants might also have based their decisions on their
perceived reliability of the summary itself. Actual AI summaries tend to be more
measured than our perturbed summaries, and it is possible that
users would react differently to a more neutral summary. However, this would have been mitigated by our randomized experiment setup, as every participant would have received the same summary, such that we could attribute any variances in answers to the experimental conditions. The summaries were perturbed from the original AI summaries, preserving the syntax and structure. Furthermore, we added the open-ended response about information sufficiency to give participants the opportunity to express this concern. Furthermore, the majority of participants (85.1\%) said they had all the information they needed,  suggesting that this did not majorly impact their decisions. 

Furthermore, we limited our survey to four products, and specifically search
products from a single platform, with a limited number of user reviews per product. This narrow selection may limit the generalizability of our findings.
Users may react to other types of products differently, especially experience
products, which customers rely upon different decision processes to purchase. Cultural products in particular often rely on word of mouth and
interpersonal connections---for instance, someone being more likely to
watch a movie when a friend whose taste aligns with yours recommends it. 
We also framed each decision as buying for a friend rather than for the
participant themselves. This was intended to reduce the influence of a
participant's own budget and demand for the product, but it also removes the
personal stake of a real purchase. Participants deciding for themselves may hold stronger prior preferences and weigh risk differently. Future research could explore how personal ties and recommendations factor into user reliance on AI generated information, as well as structuring the study as a product the user is going to buy themselves, rather than buy for a friend. 

Finally, the helpfulness ranking in Section~9 required participants to order the
three sources rather than rate each one. The mean ranks therefore reflect
relative position and not how helpful participants found each source on its own.
Collecting absolute ratings alongside the ranking would separate the two.

\section{Conclusion}

In this study, we examined the influence of AI-generated summaries relative to the user-generated information they summarize in an e-commerce setting. AI summaries shifted attitudes toward their valence, and negative summaries resulted in greater shifts than positive ones. However, participants favored user reviews: when the sources conflicted, they aligned with the reviews. In fact, the \textit{placement} of an information source was the most integral factor in its influence upon a participant---people tended to anchor on the first source they saw, and only user reviews could displace an established anchor. Moreover, participants underestimated the influence of the AI summaries, sometimes ranking their helpfulness below the user reviews even as the AI summaries influenced them more. Ultimately, as AI summaries become more common across online platforms, their placement upon the interface and the users' awareness of their effect remain important considerations for their influence. 

\bibliographystyle{ACM-Reference-Format}
\bibliography{references}

@article{liu2019framework,
  title={A framework for understanding consumer choices for others},
  author={Liu, Peggy J and Dallas, Steven K and Fitzsimons, Gavan J},
  journal={Journal of Consumer Research},
  volume={46},
  number={3},
  pages={407--434},
  year={2019},
  publisher={Oxford University Press}
}

@article{goyal2026masking,
  title={Masking or Mitigating? Deconstructing the Impact of Query Rewriting on Retriever Biases in RAG},
  author={Goyal, Agam and Mukherjee, Koyel and Saxena, Apoorv and Phukan, Anirudh and Chandrasekharan, Eshwar and Sundaram, Hari},
  journal={arXiv preprint arXiv:2604.06097},
  year={2026}
}

@article{nakayama2010has,
  title={Has the web transformed experience goods into search goods?},
  author={Nakayama, Makoto and Sutcliffe, Norma and Wan, Yun},
  journal={Electronic Markets},
  volume={20},
  number={3},
  pages={251--262},
  year={2010},
  publisher={Springer}
}

@article{Guo2020PositiveEBA,
  title={Positive emotion bias: Role of emotional content from online customer reviews in purchase decisions},
  author={Junpeng Guo and Xiaopan Wang and Yi Wu},
  journal={Journal of Retailing and Consumer Services},
  year={2020},
  volume={52},
  pages={101891},
  url={https://doi.org/10.1016/J.JRETCONSER.2019.101891}
}

@article{Chen2022TheIOA,
  title={The Impact of Online Reviews on Consumers’ Purchasing Decisions: Evidence From an Eye-Tracking Study},
  author={Tao Chen and P. Samaranayake and Xiong Cen and Mengying Qi and Y. Lan},
  journal={Frontiers in Psychology},
  year={2022},
  volume={13},
  url={https://api.semanticscholar.org/CorpusId:249476189}
}

@inproceedings{zhang2020effect,
  title={Effect of confidence and explanation on accuracy and trust calibration in AI-assisted decision making},
  author={Zhang, Yunfeng and Liao, Q Vera and Bellamy, Rachel KE},
  booktitle={Proceedings of the 2020 conference on fairness, accountability, and transparency},
  pages={295--305},
  year={2020}
}

@inproceedings{yin2019understanding,
  title={Understanding the effect of accuracy on trust in machine learning models},
  author={Yin, Ming and Wortman Vaughan, Jennifer and Wallach, Hanna},
  booktitle={Proceedings of the 2019 chi conference on human factors in computing systems},
  pages={1--12},
  year={2019}
}

@inproceedings{sharma2024generative,
  title={Generative echo chamber? effect of llm-powered search systems on diverse information seeking},
  author={Sharma, Nikhil and Liao, Q Vera and Xiao, Ziang},
  booktitle={Proceedings of the 2024 CHI Conference on Human Factors in Computing Systems},
  pages={1--17},
  year={2024}
}

@inproceedings{govers2026narratives,
  title={Narratives and Perspectives: How AI Summaries Steer Users' Opinions and Engagement on Social Media},
  author={Govers, Jarod and Sew, Cherie and Velloso, Eduardo and Kostakos, Vassilis and Goncalves, Jorge},
  booktitle={Proceedings of the 2026 CHI Conference on Human Factors in Computing Systems},
  pages={1--19},
  year={2026}
}

@inproceedings{dai2024neural,
  title={Neural retrievers are biased towards llm-generated content},
  author={Dai, Sunhao and Zhou, Yuqi and Pang, Liang and Liu, Weihao and Hu, Xiaolin and Liu, Yong and Zhang, Xiao and Wang, Gang and Xu, Jun},
  booktitle={Proceedings of the 30th ACM SIGKDD Conference on Knowledge Discovery and Data Mining},
  pages={526--537},
  year={2024}
}

@article{huang2026answer,
  title={Answer Bubbles: Information Exposure in AI-Mediated Search},
  author={Huang, Michelle and Goyal, Agam and Saha, Koustuv and Chandrasekharan, Eshwar},
  journal={arXiv preprint arXiv:2603.16138},
  year={2026}
}

@misc{Chapekis_Lieb_2025, title={Do people click on links in google ai summaries?}, url={https://www.pewresearch.org/short-reads/2025/07/22/google-users-are-less-likely-to-click-on-links-when-an-ai-summary-appears-in-the-results/}, journal={Pew Research Center}, publisher={Pew Research Center}, author={Chapekis, Athena and Lieb, Anna}, year={2025}, month={Jul}}

@article{klingbeil2024trust,
  title={Trust and reliance on AI—An experimental study on the extent and costs of overreliance on AI},
  author={Klingbeil, Artur and Gr{\"u}tzner, Cassandra and Schreck, Philipp},
  journal={Computers in Human Behavior},
  volume={160},
  pages={108352},
  year={2024},
  publisher={Elsevier}
}

@misc{Vaughn_Schermerhorn_2023, title={How Amazon continues to improve the customer reviews experience with Generative AI}, url={https://www.aboutamazon.com/news/amazon-ai/amazon-improves-customer-reviews-with-generative-ai}, journal={Amazon News}, publisher={US About Amazon}, author={Vaughn Schermerhorn, Director}, year={2023}, month={Aug}}

@misc{Levine_2024, title={Here’s how Amazon’s AI-generated review highlights help you make better shopping decisions}, url={https://www.aboutamazon.com/news/retail/amazon-ai-generated-review-highlights}, journal={Amazon News}, publisher={US About Amazon}, author={Levine, Ivy}, year={2024}, month={Jul}}

@article{lackermair2013importance,
  title={Importance of online product reviews from a consumer’s perspective},
  author={Lackermair, Georg and Kailer, Daniel and Kanmaz, Kenan},
  journal={Advances in economics and business},
  volume={1},
  number={1},
  pages={1--5},
  year={2013}
}

@misc{Palmer_2023, title={Amazon is using Generative A.I. to summarize product reviews}, url={https://www.cnbc.com/2023/06/12/amazon-is-using-generative-ai-to-summarize-product-reviews.html}, journal={CNBC}, publisher={CNBC}, author={Palmer, Annie}, year={2023}, month={Jun}}

@article{loureiro2020norms,
  title={Norms for 150 consumer products: Perceived complexity, quality objectivity, material/experiential nature, perceived price, familiarity and attitude},
  author={Loureiro, Filipe and Garcia-Marques, Teresa and Wegener, Duane T},
  journal={Plos one},
  volume={15},
  number={9},
  pages={e0238848},
  year={2020},
  publisher={Public Library of Science San Francisco, CA USA}
}

@article{kyto2019comparison,
  title={Comparison of explicit vs. implicit measurements in predicting food purchases},
  author={Kyt{\"o}, Elina and Bult, Harold and Aarts, Esther and Wegman, Joost and Ruijschop, Rianne MAJ and Mustonen, Sari},
  journal={Food Quality and Preference},
  volume={78},
  pages={103733},
  year={2019},
  publisher={Elsevier}
}

@misc{li2024utilizinghumanbehaviormodeling,
      title={Utilizing Human Behavior Modeling to Manipulate Explanations in AI-Assisted Decision Making: The Good, the Bad, and the Scary}, 
      author={Zhuoyan Li and Ming Yin},
      year={2024},
      eprint={2411.10461},
      archivePrefix={arXiv},
      primaryClass={cs.HC},
      url={https://arxiv.org/abs/2411.10461}, 
}

@article{nelson1970information,
  title={Information and consumer behavior},
  author={Nelson, Phillip},
  journal={Journal of political economy},
  volume={78},
  number={2},
  pages={311--329},
  year={1970},
  publisher={The University of Chicago Press}
}

@inproceedings{airbnb,
author = {Qiu, Will and Parigi, Paolo and Abrahao, Bruno},
year = {2018},
month = {04},
pages = {1-11},
title = {More Stars or More Reviews?},
doi = {10.1145/3173574.3173727}
}

@inproceedings{li2026guided,
  title={Guided Reflection in AI-Assisted Decision-Making: Effects on AI Overreliance and Decision Accuracy},
  author={Li, Shanshan and Li, Jingwei and Li, Huiran and Zhu, Hongwei and Li, Xitong},
  booktitle={Proceedings of the 2026 CHI Conference on Human Factors in Computing Systems},
  pages={1--19},
  year={2026}
}

@article{nelson1974advertising,
  title={Advertising as information},
  author={Nelson, Phillip},
  journal={Journal of political economy},
  volume={82},
  number={4},
  pages={729--754},
  year={1974},
  publisher={The University of Chicago Press}
}

@article{salvi2026commercial,
  title={Commercial Persuasion in AI-Mediated Conversations},
  author={Salvi, Francesco and Cuevas, Alejandro and Ribeiro, Manoel Horta},
  journal={arXiv preprint arXiv:2604.04263},
  year={2026}
}

@article{cui2012effect,
  title={The effect of online consumer reviews on new product sales},
  author={Cui, Geng and Lui, Hon-Kwong and Guo, Xiaoning},
  journal={International journal of electronic commerce},
  volume={17},
  number={1},
  pages={39--58},
  year={2012},
  publisher={Taylor \& Francis}
}

@article{castelo2019task,
  title={Task-dependent algorithm aversion},
  author={Castelo, Noah and Bos, Maarten W and Lehmann, Donald R},
  journal={Journal of marketing research},
  volume={56},
  number={5},
  pages={809--825},
  year={2019},
  publisher={SAGE Publications Sage CA: Los Angeles, CA}
}

@inproceedings{li2025text,
  title={From text to trust: empowering ai-assisted decision making with adaptive LLM-powered analysis},
  author={Li, Zhuoyan and Zhu, Hangxiao and Lu, Zhuoran and Xiao, Ziang and Yin, Ming},
  booktitle={Proceedings of the 2025 CHI Conference on Human Factors in Computing Systems},
  pages={1--18},
  year={2025}
}

@article{selfverification,
author = {Schmalz, Marc and Carter, Michelle and Lee, Jin Ha},
title = {It's Not You, It's Me: Identity, Self-Verification, and Amazon Reviews},
year = {2018},
issue_date = {May 2018},
publisher = {Association for Computing Machinery},
address = {New York, NY, USA},
volume = {49},
number = {2},
issn = {0095-0033},
url = {https://doi.org/10.1145/3229335.3229341},
doi = {10.1145/3229335.3229341},
journal = {SIGMIS Database},
month = may,
pages = {79–92},
numpages = {14}
}

@article{informationaccess,
author = {Shah, Chirag and Bender, Emily M.},
title = {Envisioning Information Access Systems: What Makes for Good Tools and a Healthy Web?},
year = {2024},
issue_date = {August 2024},
publisher = {Association for Computing Machinery},
address = {New York, NY, USA},
volume = {18},
number = {3},
issn = {1559-1131},
url = {https://doi.org/10.1145/3649468},
doi = {10.1145/3649468},
journal = {ACM Trans. Web},
month = apr,
articleno = {33},
numpages = {24}
}

@inproceedings{kelly2025understanding,
  title={Understanding Gender Bias in AI-Generated Product Descriptions},
  author={Kelly, Markelle and Tahaei, Mohammad and Smyth, Padhraic and Wilcox, Lauren},
  booktitle={Proceedings of the 2025 ACM Conference on Fairness, Accountability, and Transparency},
  pages={2587--2615},
  year={2025}
}

@article{dellarocas2003digitization,
  title={The digitization of word of mouth: Promise and challenges of online feedback mechanisms},
  author={Dellarocas, Chrysanthos},
  journal={Management science},
  volume={49},
  number={10},
  pages={1407--1424},
  year={2003},
  publisher={INFORMS}
}

@article{chen2008all,
  title={All reviews are not created equal: The disaggregate impact of reviews and reviewers at amazon. com},
  author={Chen, Pei-Yu and Dhanasobhon, Samita and Smith, Michael D},
  journal={Com (May 2008)},
  year={2008}
}

@article{cragg1971some,
  title={Some statistical models for limited dependent variables with application to the demand for durable goods},
  author={Cragg, John G},
  journal={Econometrica: journal of the Econometric Society},
  pages={829--844},
  year={1971},
  publisher={JSTOR}
}

@article{mullahy1986specification,
  title={Specification and testing of some modified count data models},
  author={Mullahy, John},
  journal={Journal of econometrics},
  volume={33},
  number={3},
  pages={341--365},
  year={1986},
  publisher={Elsevier}
}

@article{park2024generative,
  title={Generative agent simulations of 1,000 people},
  author={Park, Joon Sung and Zou, Carolyn Q and Shaw, Aaron and Hill, Benjamin Mako and Cai, Carrie and Morris, Meredith Ringel and Willer, Robb and Liang, Percy and Bernstein, Michael S},
  journal={arXiv preprint arXiv:2411.10109},
  volume={52},
  year={2024}
}

@article{yao2025generative,
  title={Generative ai for simulating real world dynamics applications and challenges},
  author={Yao, Chenxi and Zhan, Qishi and Cao, Zeyu and Lin, Yangfan and Li, Danping and Shao, Yuanxun and Wang, Lin and Wang, Zhao and Zhang, Jiaqing and Zhang, Yunfei and others},
  year={2025},
  publisher={TechRxiv}
}

@article{ghaffarzadegan2024generative,
  title={Generative agent-based modeling: an introduction and tutorial},
  author={Ghaffarzadegan, Navid and Majumdar, Aritra and Williams, Ross and Hosseinichimeh, Niyousha},
  journal={System Dynamics Review},
  volume={40},
  number={1},
  pages={e1761},
  year={2024},
  publisher={Wiley Online Library}
}

@inproceedings{park2023generative,
  title={Generative agents: Interactive simulacra of human behavior},
  author={Park, Joon Sung and O'Brien, Joseph and Cai, Carrie Jun and Morris, Meredith Ringel and Liang, Percy and Bernstein, Michael S},
  booktitle={Proceedings of the 36th annual acm symposium on user interface software and technology},
  pages={1--22},
  year={2023}
}

@article{serapio2023personality,
  title={Personality traits in large language models},
  author={Serapio-Garc{\'\i}a, Greg and Safdari, Mustafa and Crepy, Cl{\'e}ment and Sun, Luning and Fitz, Stephen and Romero, Peter and Abdulhai, Marwa and Faust, Aleksandra and Matari{\'c}, Maja},
  journal={arXiv preprint arXiv:2307.00184},
  year={2023}
}

@article{williams2023epidemic,
  title={Epidemic modeling with generative agents},
  author={Williams, Ross and Hosseinichimeh, Niyousha and Majumdar, Aritra and Ghaffarzadegan, Navid},
  journal={arXiv preprint arXiv:2307.04986},
  year={2023}
}

@inproceedings{cheng2023marked,
  title={Marked personas: Using natural language prompts to measure stereotypes in language models},
  author={Cheng, Myra and Durmus, Esin and Jurafsky, Dan},
  booktitle={Proceedings of the 61st Annual Meeting of the Association for Computational Linguistics (Volume 1: Long Papers)},
  pages={1504--1532},
  year={2023}
}

@article{zou2024can,
  title={Can llm" self-report"?: Evaluating the validity of self-report scales in measuring personality design in llm-based chatbots},
  author={Zou, Huiqi and Wang, Pengda and Yan, Zihan and Sun, Tianjun and Xiao, Ziang},
  journal={arXiv preprint arXiv:2412.00207},
  year={2024}
}

@article{buccinca2021trust,
  title={To trust or to think: cognitive forcing functions can reduce overreliance on AI in AI-assisted decision-making},
  author={Bu{\c{c}}inca, Zana and Malaya, Maja Barbara and Gajos, Krzysztof Z},
  journal={Proceedings of the ACM on Human-computer Interaction},
  volume={5},
  number={CSCW1},
  pages={1--21},
  year={2021},
  publisher={ACM New York, NY, USA}
}

@inproceedings{wang2024incharacter,
  title={Incharacter: Evaluating personality fidelity in role-playing agents through psychological interviews},
  author={Wang, Xintao and Xiao, Yunze and Huang, Jen-tse and Yuan, Siyu and Xu, Rui and Guo, Haoran and Tu, Quan and Fei, Yaying and Leng, Ziang and Wang, Wei and others},
  booktitle={Proceedings of the 62nd annual meeting of the association for computational linguistics (volume 1: Long papers)},
  pages={1840--1873},
  year={2024}
}

@article{tu2023characterchat,
  title={Characterchat: Learning towards conversational ai with personalized social support},
  author={Tu, Quan and Chen, Chuanqi and Li, Jinpeng and Li, Yanran and Shang, Shuo and Zhao, Dongyan and Wang, Ran and Yan, Rui},
  journal={arXiv preprint arXiv:2308.10278},
  year={2023}
}

\appendix

\section{Demographic Information}

\begin{table}[ht]
\centering\sffamily
\caption{Age distribution}\Description{Age distribution of the 278 participants. Ages 35–44 are the largest group with 73 participants, or 26.26 percent, followed by ages 25–34 with 71, or 25.54 percent; ages 45–54 with 61, or 21.94 percent; ages 18–24 with 48, or 17.27 percent; ages 55–64 with 18, or 6.47 percent; and ages 65 or older with 7, or 2.52 percent.}
\label{tab:age}
\begin{tabular}{lrr}
\rowcolor{blue!10}\textbf{Age group} & \textit{n} & \textit{\%} \\
\midrule
    \rowcolor{gray!10}35–44 & 73 & 26.26\% \\
    25–34 & 71 & 25.54\% \\
    \rowcolor{gray!10}45–54 & 61 & 21.94\% \\
    18–24 & 48 & 17.27\% \\
    \rowcolor{gray!10}55–64 & 18 & 6.47\% \\
    65 or older & 7 & 2.52\% \\

    \midrule
    \textbf{Total} & \textbf{278} & \textbf{100.00\%} \\
\bottomrule
\end{tabular}
\end{table}

\begin{table}[ht]
\centering\sffamily
\caption{Sex distribution}\Description{Sex distribution of the 278 participants. There are 168 female participants, representing 60.43 percent of the sample; 109 male participants, representing 39.21 percent; and one participant, or 0.36 percent, who preferred not to state their sex.}
\label{tab:sex}
\begin{tabular}{lrr}
\rowcolor{blue!10}\textbf{Sex} & \textit{n} & \textit{\%} \\
\midrule
    \rowcolor{gray!10}Female & 168 & 60.43\% \\
    Male & 109 & 39.21\% \\
    \rowcolor{gray!10}Prefer not to say & 1 & 0.36\% \\
\midrule
    \textbf{Total} & \textbf{278} & \textbf{100.00\%} \\
\bottomrule
\end{tabular}
\end{table}

\begin{table}[H]
\centering\sffamily
\caption{Ethnicity distribution}\Description{Simplified ethnicity distribution of the 278 participants. The sample includes 174 White participants, or 62.59 percent; 46 Black participants, or 16.55 percent; 28 Asian participants, or 10.07 percent; 17 participants categorized as Mixed, or 6.12 percent; and 13 participants categorized as Other, or 4.68 percent.}
\label{tab:ethnicity}
\begin{tabular}{lrr}
\rowcolor{blue!10}\textbf{Ethnicity simplified} & \textit{n} & \textit{\%} \\
\midrule
    \rowcolor{gray!10}White & 174 & 62.59\% \\
    Black & 46 & 16.55\% \\
    \rowcolor{gray!10}Asian & 28 & 10.07\% \\
    Mixed & 17 & 6.12\% \\
    \rowcolor{gray!10}Other & 13 & 4.68\% \\
\midrule
    \textbf{Total} & \textbf{278} & \textbf{100.00\%} \\
\bottomrule
\end{tabular}
\end{table}

\section{User Reviews and Perturbed Summaries}
\label{Appendix-reviews}

\begin{figure*}[ht]
    \centering
    \includegraphics[width=0.65\textwidth]{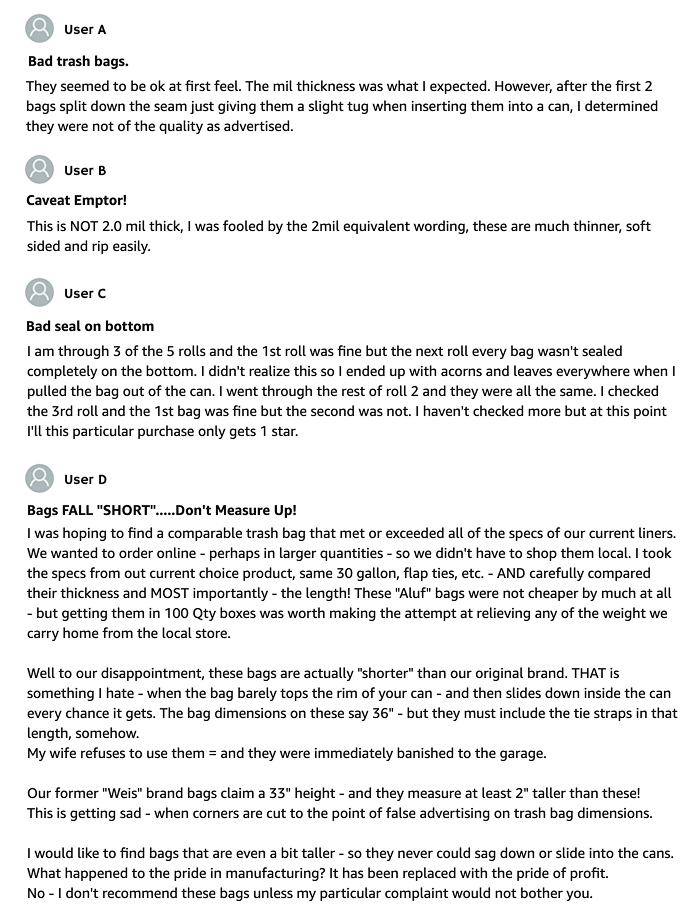}
    \caption{Product 1 Negative User Reviews}
    \Description{Four negative customer reviews of trash bags.}
    \label{fig:p1-neg-user}
\end{figure*}

\begin{figure*}[ht]
    \centering
    \includegraphics[width=0.65\textwidth]{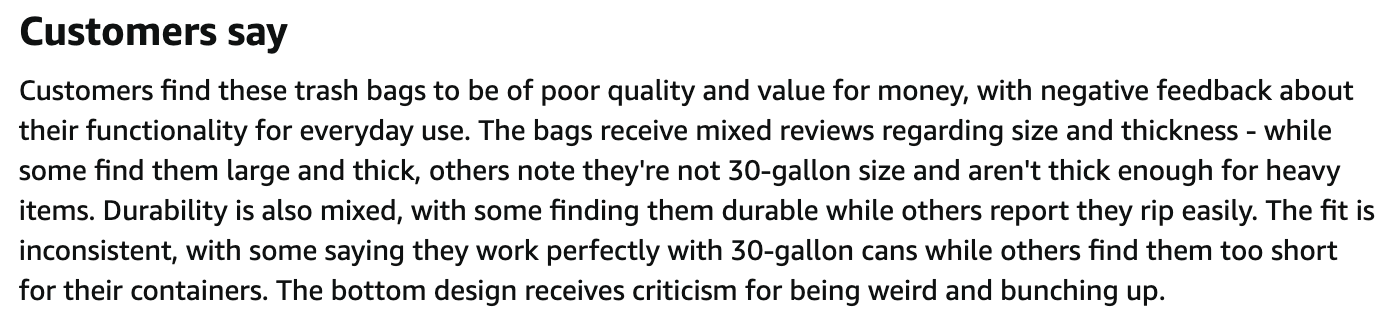}
    \caption{Product 1 Negative AI Summary}
    \Description{Negative AI summary of customer opinions about the trash bags.}
    \label{fig:p1-neg-ai}
\end{figure*}

\begin{figure*}[ht]
    \centering
    \includegraphics[width=0.65\textwidth]{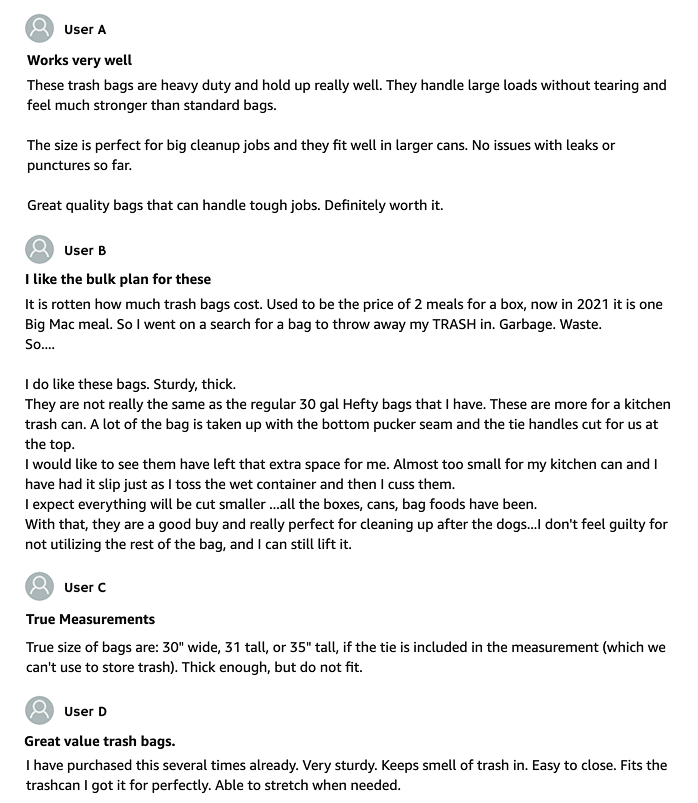}
    \caption{Product 1 Positive User Reviews}
    \Description{Four positive customer reviews of trash bags.}
    \label{fig:p1-pos-user}
\end{figure*}

\begin{figure*}[ht]
    \centering
    \includegraphics[width=0.65\textwidth]{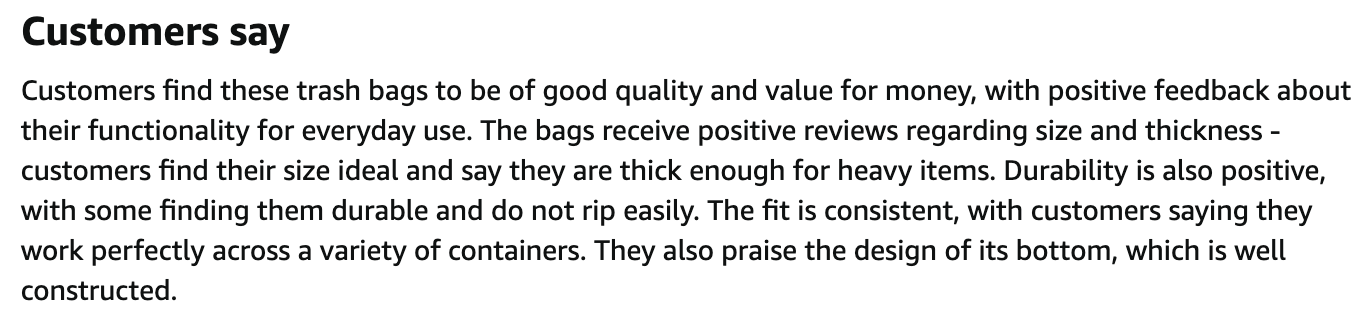}
    \caption{Product 1 Positive AI Summary}
    \Description{Positive AI summary of customer opinions about the trash bags.}
    \label{fig:p1-pos-ai}
\end{figure*}

\begin{figure*}[ht]
    \centering
    \includegraphics[width=0.65\textwidth]{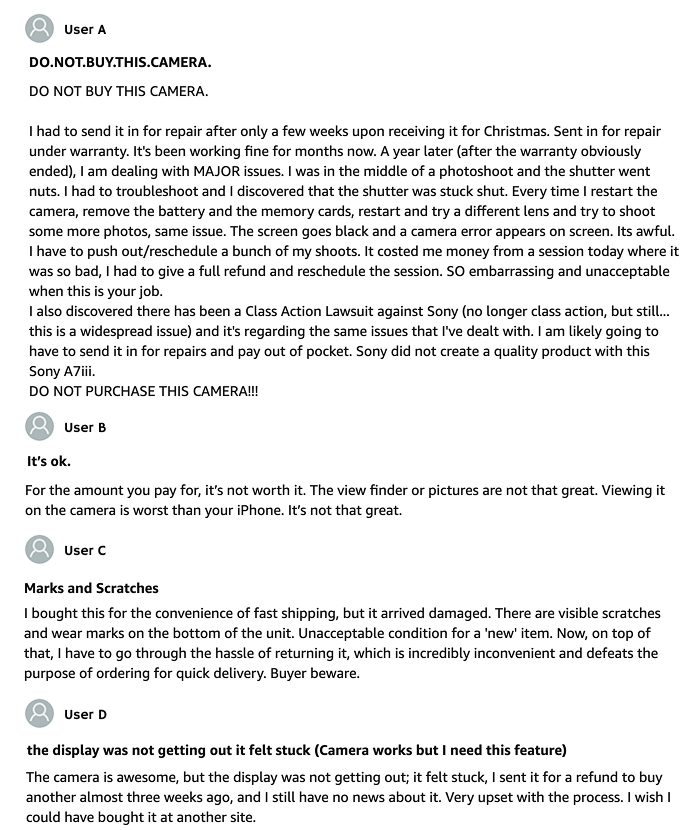}
    \caption{Product 2 Negative User Reviews}
    \Description{Four negative customer reviews of a camera.}
    \label{fig:p2-neg-user}
\end{figure*}

\begin{figure*}[ht]
    \centering
    \includegraphics[width=0.65\textwidth]{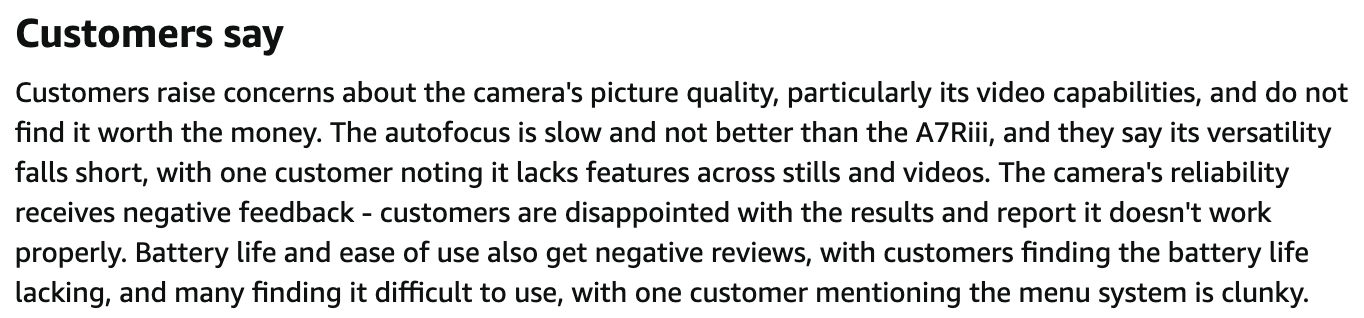}
    \caption{Product 2 Negative AI Summary}
    \Description{Negative AI summary of customer opinions about the camera.}
    \label{fig:p2-neg-ai}
\end{figure*}

\begin{figure*}[ht]
    \centering
    \includegraphics[width=0.65\textwidth]{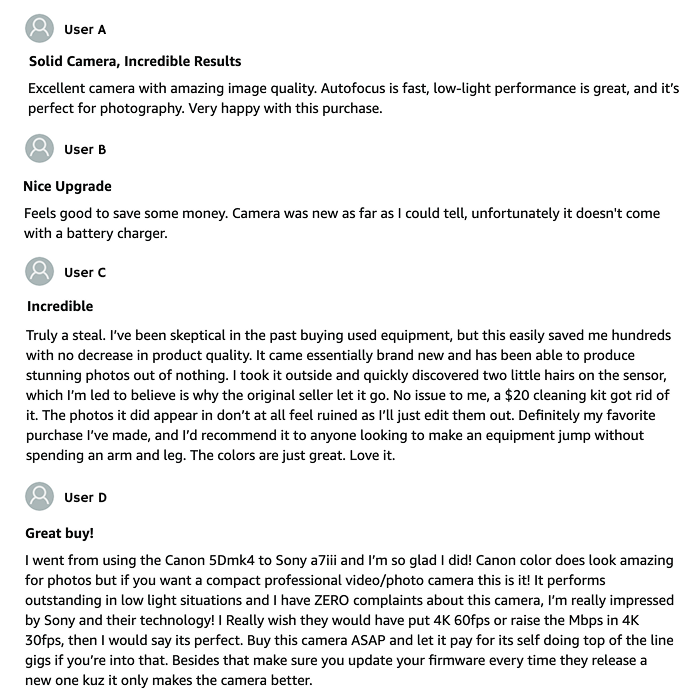}
    \caption{Product 2 Positive User Reviews}
    \Description{Four positive customer reviews of a camera.}
    \label{fig:p2-pos-user}
\end{figure*}

\begin{figure*}[ht]
    \centering
    \includegraphics[width=0.65\textwidth]{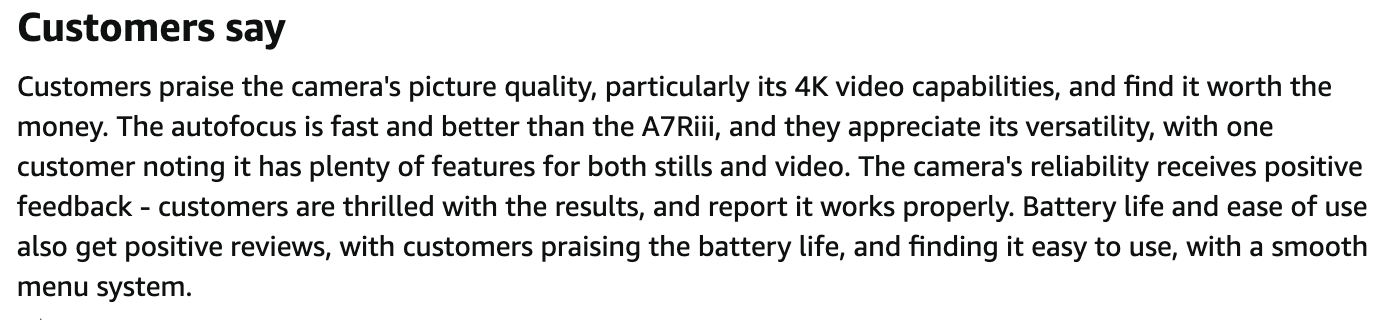}
    \caption{Product 2 Positive AI Summary}
    \Description{Positive AI summary of customer opinions about the camera.}
    \label{fig:p2-pos-ai}
\end{figure*}

\begin{figure*}[ht]
    \centering
    \includegraphics[width=0.65\textwidth]{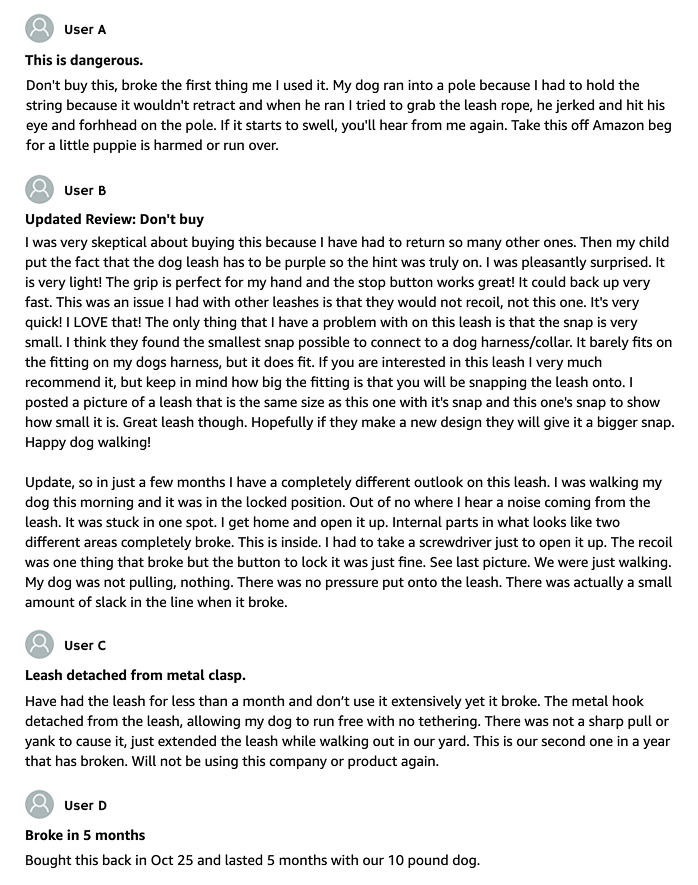}
    \caption{Product 3 Negative User Reviews}\Description{Four negative customer reviews of a retractable dog leash.}
    \label{fig:p3-neg-user}
\end{figure*}

\begin{figure*}[ht]
    \centering
    \includegraphics[width=0.65\textwidth]{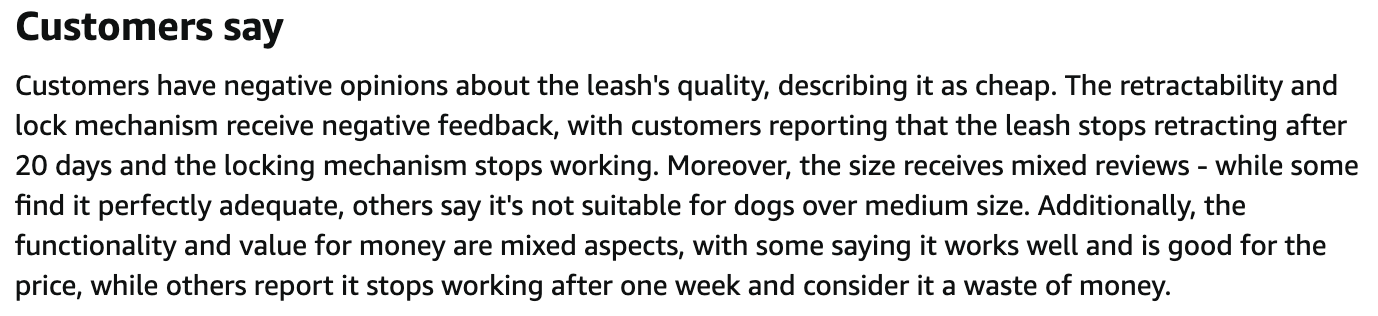}
    \caption{Product 3 Negative AI Summary}
    \Description{Negative AI summary of customer opinions about the retractable dog leash.}
    \label{fig:p3-neg-ai}
\end{figure*}

\begin{figure*}[ht]
    \centering
    \includegraphics[width=0.65\textwidth]{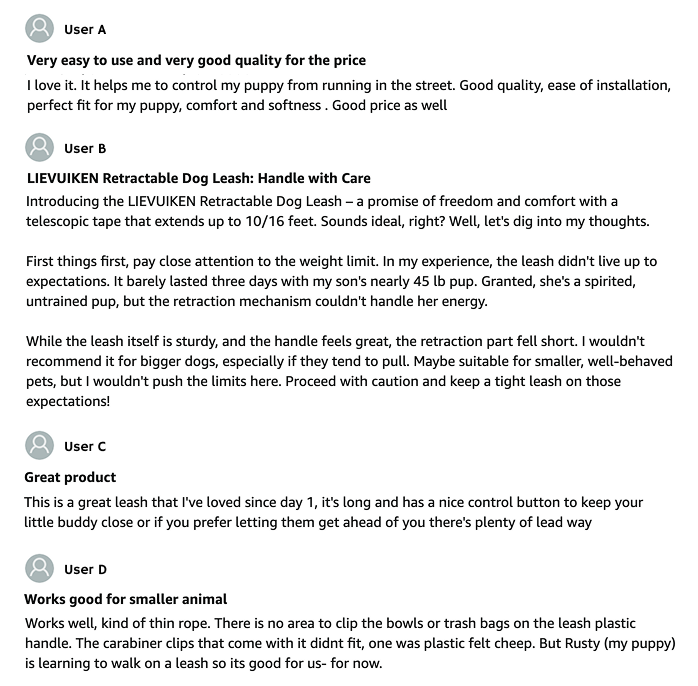}
    \caption{Product 3 Positive User Reviews}\Description{Four positive customer reviews of a retractable dog leash.}
    \label{fig:p3-pos-user}
\end{figure*}

\begin{figure*}[ht]
    \centering
    \includegraphics[width=0.65\textwidth]{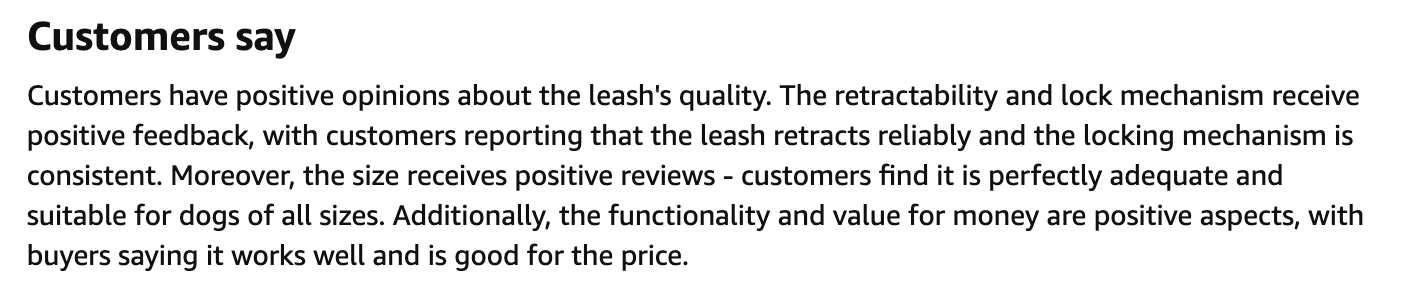}
    \caption{Product 3 Positive AI Summary}
    \Description{Positive AI summary of customer opinions about the dog leash.}
    \label{fig:p3-pos-ai}
\end{figure*}

\begin{figure*}[!htbp]
    \centering
    \includegraphics[width=0.5\textwidth]{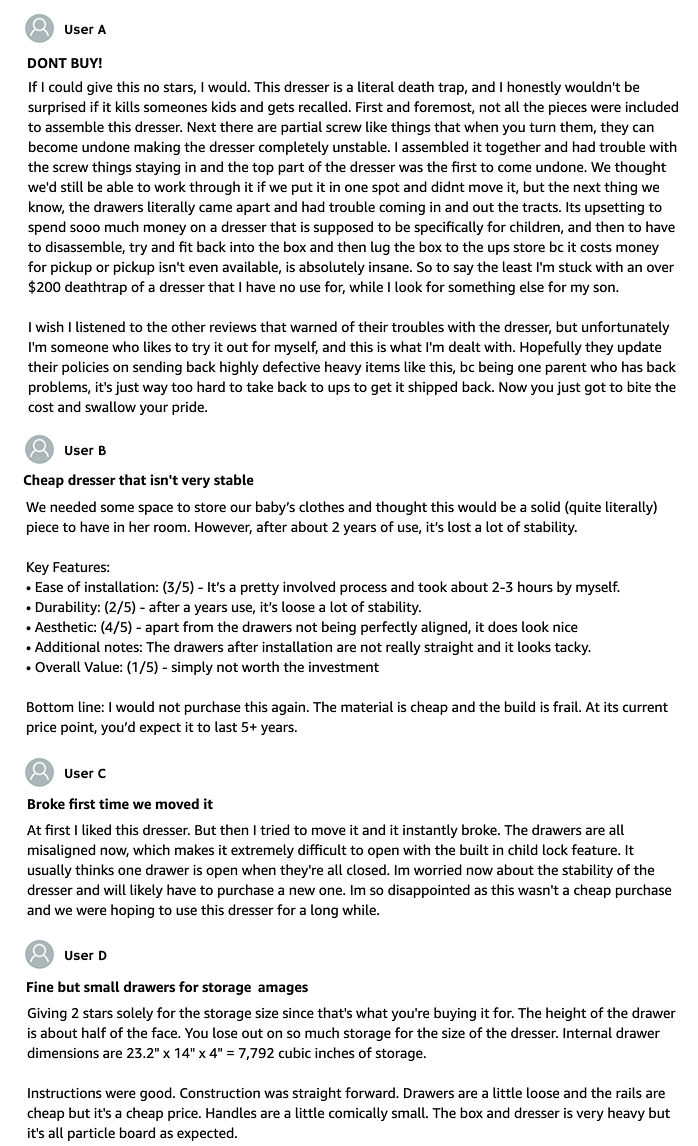}
    \caption{Product 4 Negative User Reviews}\Description{Four negative or mixed customer reviews of a changing-table dresser.}
    \label{fig:p4-neg-user}
\end{figure*}

\begin{figure*}[!htbp]
    \centering
    \includegraphics[width=0.65\textwidth]{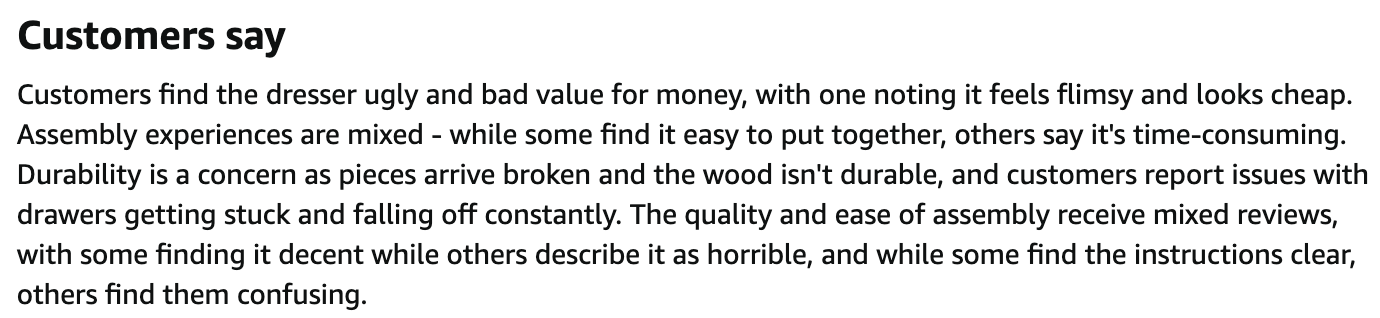}
    \caption{Product 4 Negative AI Summary}
    \Description{Negative AI summary of customer opinions about the changing-table dresser.}
    \label{fig:p4-neg-ai}
\end{figure*}

\begin{figure*}[!htbp]
    \centering
    \includegraphics[width=0.65\textwidth]{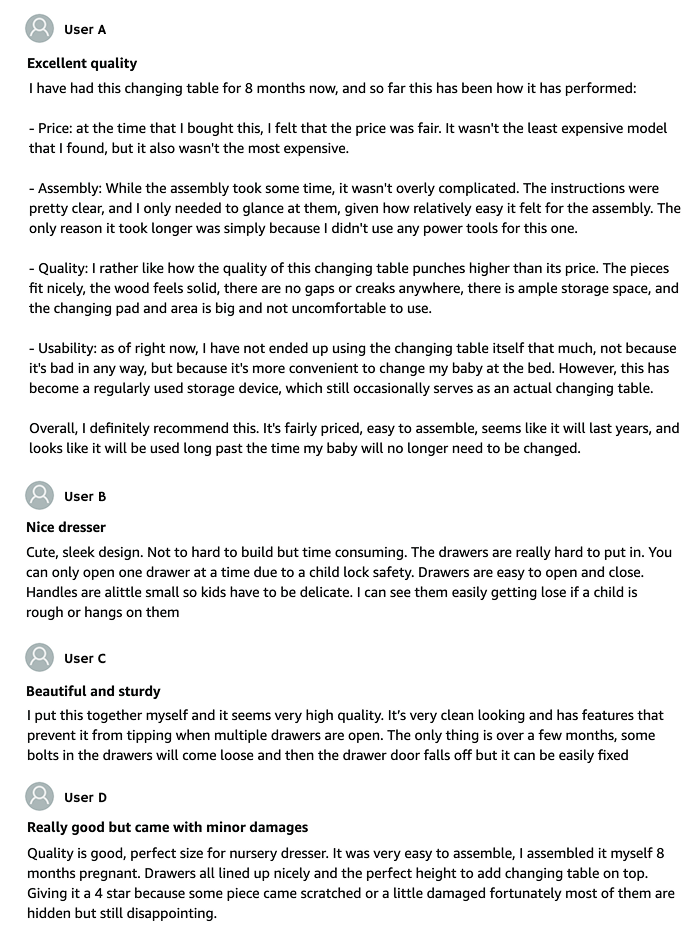}
    \caption{Product 4 Positive User Reviews}\Description{Four positive customer reviews of a changing-table dresser.}
    \label{fig:p4-pos-user}
\end{figure*}
\begin{figure*}[!htbp]
    \centering
    \includegraphics[width=0.65\textwidth]{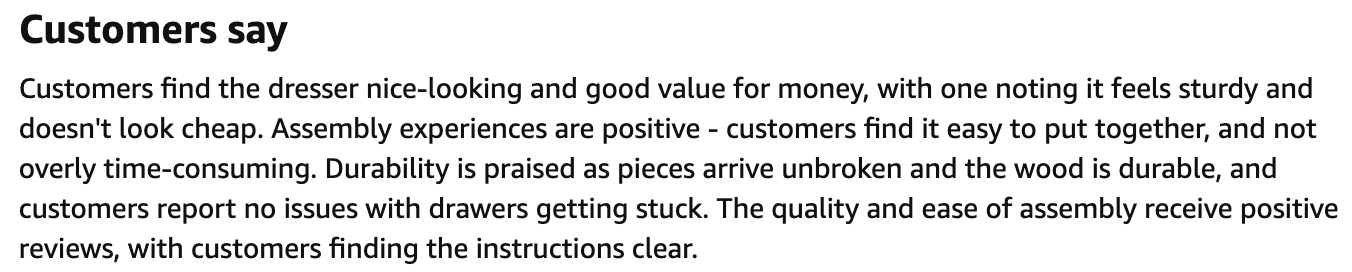}
    \caption{Product 4 Positive AI Summary}
    \Description{Positive AI summary of customer opinions about the changing-table dresser.}
    \label{fig:p4-pos-ai}
\end{figure*}

\end{document}